\documentclass[twocolumn, showpacs,preprintnumbers,amsmath,amssymb]{revtex4}
\usepackage{enumitem}
\usepackage{amssymb}
\usepackage{xcolor}
\usepackage{graphicx}
\usepackage{dcolumn}
\usepackage{bm}
\usepackage{multirow}
\usepackage{booktabs}
\usepackage{longtable}
\usepackage{natbib}
\usepackage{lscape}
\usepackage{soul}
\usepackage{amssymb}
\usepackage{amsmath}	
\usepackage{color}
\usepackage{cases} 

\usepackage{array}
\newcommand{\Kmat}{\mbox{\boldmath $\mathcal{K}$}}

\newcommand{\V}{\mbox{\boldmath $V$}}

\newcommand{\Omat}{{\bf O}}

\newcommand{\MV}{\mathcal{V}}

\newcommand{\Vel}{\ensuremath{V_\mathrm{el}}}
\newcommand{\SH}{\ensuremath{\mathrm{SH}}}
\newcommand{\SHplus}{\ensuremath{\mathrm{SH}^+}}
\newcommand{\etal}{{\it et~al.}}
\newcommand{\Hdia}{\ensuremath{\mathbf{H}_\mathrm{dia}}}
\newcommand{\Hel}{\ensuremath{H_\mathrm{el}}}

\begin{document}
\preprint{APS/123-QED}

\title{A theoretical study of the dissociative recombination of SH$^+$ with electrons through the $^2\Pi$ states of SH}
\author{D.O. Kashinski$^{1}$}\email{david.kashinski@usma.edu}
\author{D. Talbi$^{2}$}\email{dahbia.talbi@umontpellier.fr}
\author{A.P. Hickman$^{3}$}\email{aph2@lehigh.edu}
\author{O.E. Di Nallo$^{1}$}
\author{F. Colboc$^{4}$}
\author{K. Chakrabarti$^{5}$}
\author{I. F. Schneider$^{4}$}\email[]{ioan.schneider@univ-lehavre.fr}
\author{J. Zs Mezei$^{4,6,7}$}
\affiliation{$^{1}$Dept. of Physics and Nuclear Engineering, United States Military Academy, West Point, NY, 10996, USA}%
\affiliation{$^{2}$LUPM CNRS$-$Universit{\'{e}} de Montpellier, 34095 Montpellier, France}
\affiliation{$^{3}$Dept. oof Physics, Lehigh University, Bethlehem, PA 18015, USA}%
\affiliation{$^{4}$LOMC CNRS$-$Universit{\'{e}} du Havre$-$Normandie Universit{\'{e}}, 76058 Le Havre, France}
\affiliation{$^{5}$Dept. of Mathematics, Scottish Church College, Calcutta 700 006, India}
\affiliation{$^{6}$LAC, CNRS$-$Universit\'e Paris-Sud$-$ENS Cachan$-$Universit\'e Paris-Saclay, 91405 Orsay, France}%
\affiliation{$^{7}$LSPM, CNRS$-$Universit\'e Paris 13$-$USPC, 93430 Villetaneuse, France}%
\date{\today}

\begin{abstract}
A quantitative theoretical study of the dissociative recombination of \SHplus\ with electrons has been carried out. Multireference, configuration interaction calculations were used to determine accurate potential energy curves for \SHplus\ and \SH. The block diagonalization method was used to disentangle strongly interacting \SH\ valence and Rydberg states and to construct a diabatic Hamiltonian whose diagonal matrix elements  provide the diabatic potential energy curves. The off-diagonal elements  are related to the electronic valence-Rydberg couplings. Cross sections and rate coefficients for the dissociative recombination reaction were calculated with a step-wise version of the multichannel quantum defect theory, using the molecular data provided by the block diagonalization method. The calculated rates are compared with the most recent measurements  performed on the TSR ion storage ring in Heidelberg, Germany.
\end{abstract}


\maketitle

\section{Introduction}

The 526 GHz $N_J = 1_2$--$0_1$ transition of the \SHplus\ radical (sulfoniumylidene or sulfanylium) was first detected in emission in W3
IRS 5, a prototypical region of high-mass star formation \cite{Benz2010}.
\SHplus\ has also been observed in absorption from the quartet of its
ground-state hyperfine structure near 683\,GHz towards Sagitarius B2 and
found to be ubiquitous in the diffuse interstellar medium \cite{Menten2011}.
The 526\,GHz transition has also been detected in absorption in the
diffuse interstellar medium towards various distant star-forming regions
\cite{Godard2012}. \SHplus\ has also been observed in emission in the Orion
Bar, a typical high UV-illumination warm and dense photon-dominated region
(PDR), through the same 526 and 683 GHz lines \cite{2Nagy2013}.

The chemistry of \SHplus\ in the interstellar medium (ISM) is not yet well
understood \cite{Millar1986}. Its most likely formation channel, the
reaction of atomic sulfur S$^+$ with H$_2$, is highly endothermic by 0.86~eV
(9860K). For the \SHplus\ reaction product to be formed in the diffuse ISM at
sufficient abundance to be observed, it has been suggested that the
endothermicity related to its formation must be overcome by turbulent
dissipation, shocks, or shears. For that reason \SHplus\ has been proposed as
an important turbulence probe in the diffuse ISM \cite{Godard2012,Godard2014}. In photon dominated regions where H$_2$ is vibrationally excited, it has been suggested that it is this excess of vibrational energy that allows for the formation of \SHplus\ from S$^+$
and H$_2$ \cite{Zanchet2013}.

\begin{figure}
\centering
\includegraphics[width=0.9\columnwidth]{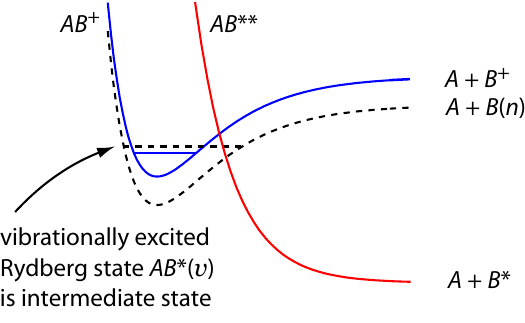}
\caption{\label{fig:DRexplain2} Schematic potential  energy curves for a
  molecular ion AB$^+$ and the dissociating neutral state relevant to  DR.
A representative Rydberg curve is shown by the dashed line.}
\end{figure}

It is obvious from the above examples that the analysis of the \SHplus\ ion in
different environments of the ISM could provide unique physical and chemical
constraints on models that describe these environments. Therefore the \SHplus\ chemistry needs to be known in detail. This is not yet the case. For instance, \SHplus\ does not react quickly with H$_2$ \cite{Millar1986}, the most abundant interstellar molecule, because the reaction is endothermic; therefore \SHplus\ is not severely depleted by this process. But it is thought
that \SHplus\ is efficiently destroyed through dissociative recombination (DR) with electrons,
\begin{equation}\label{eq:reac-dr}
 \mathrm{SH}^+  + e^- \rightarrow   \mathrm{S} + \mathrm{H},
\end{equation}
and a rate constant of $10^{-6}\,\mathrm{cm^3\,s^{-1}}$ is assigned to this reaction in astrochemical databases \cite{kida}. However, at the time we started the present work, no theoretical or experimental study was available that could confirm or rule out the efficient destruction of \SHplus\ by DR, which would of course greatly influence the  abundance of \SHplus\ in the ISM. There was therefore a strong demand from the astrophysical community for an investigation of this process.  We have undertaken this task using the tools of theoretical chemistry and of collision dynamics. Such a task requires calculating the potential energy curves (PECs) governing the DR of \SHplus, namely, those of the initial molecular ion and, for the corresponding neutral, the PECs of the $^{2,4}\Sigma$, $^{2,4}\Pi$, and $^{2,4}\Delta$ Rydberg  and  dissociating autoionizing states. Typical behavior for these curves is illustrated schematically in Fig.~\ref{fig:DRexplain2}. For reasons discussed in Section~\ref{sec:SOCIcalcs}, the calculations reported here consider only the $^2\Pi$ states of SH. The validity of this simplification will be assessed by comparing the calculations with very recent results  \cite{becker2015} from measurements at the TSR ion storage ring in Heidelberg, Germany.

The literature reports a large variety of theoretical studies on the SH$^+$ ion (Khadri \etal\ \cite{Khadri2006} and reference therein) as well as on the SH radical. The most extensive study on the excited electronic states of SH is from Bruna and Hirsch \cite{Bruna1987}, which completes previous works by Meyer \etal\ \cite{Meyer1975} and Hirst and Guest~\cite{Hirst1982}. Bruna and Hirsch \cite{Bruna1987} have performed Multireference, Double excitation Configuration Interaction (MRD-CI) calculations on SH in its low-lying valence states as well as on its first Rydberg terms and have been able to give a complete reassignment of the electronic structures of the excited states of SH determined experimentally by Morrow \cite{Morrow1966}. Park and Sun~\cite{park1992}, using effective valence shell Hamiltonian calculations, have confirmed the identification by Bruna and Hirsch \cite{Bruna1987} of the $2\, ^2\Sigma^-$ and $3\, ^2\Sigma^-$ SH excited states as Rydberg states, but while Bruna and Hirsch \cite{Bruna1987} have attributed a valence character at the vertical region for the $1\,^{2}\Sigma^{-}$ and $2\,^{2}\Sigma^{+}$ of SH, Park and Sun~\cite{park1992} have identified these states as Rydberg. It is known that the  $X\,^2\Pi$ and $A\,^2\Sigma$ states of SH are of valence character. The latter has been extensively studied because of its role in the photodissociation of SH (Rose \etal~\cite{Rose2009} and reference therein).

While rather extensive, none of the studies cited above provide the curves needed for understanding  the DR of SH$^+$. In the present paper we report extensive Multireference Configuration Interaction (MRCI) calculations of the ground, excited valence, and Rydberg PECs of SH and the ground and first excited PECs of \SHplus, calculated at the same level of theory. Using the Block diagonalization method~\cite{Dom93,Spi00,Hic05,Hic11,Kashinski2012} (described in section \ref{sec:BLOCKD}) we extracted the diabatic PECs needed to understand and to quantify the DR of SH$^+$, namely the curves illustrated in Fig.~\ref{fig:DRexplain2}. Multichannel Quantum Defect Theory (MQDT)~\cite{jungen96} has been used together with the results from the MRCI calculations to determine DR cross sections and rate constants as a function of collision energies relevant for the interstellar medium.

This paper is organized as follows.  Section \ref{sec:ElecStr} describes the electronic
structure calculations, and Section \ref{sec:BLOCKD} shows how they were used to determine
the necessary PECs and coupling matrix elements.  Section \ref{sec:mqdt} presents the calculations of the DR cross sections and rate constants and compares them with recent experiments. Section \ref{sec:conclude} contains concluding remarks.

\section{Electronic Structure Calculations} \label{sec:ElecStr}

\subsection{General Considerations} \label{sec:Generalities}

Calculating appropriate PECs to treat DR  is difficult because the process involves dissociating autoionizing states that are embedded in the continuum of electron--ion scattering states.
These states are highly excited and mix strongly with other excited states, leading to many avoided crossings. Previous work by ourselves \cite{Spi00,Hic05,Hic11,Kashinski2012}
and others \cite{pacher1988,Dom93} has shown that the block diagonalization method \cite{pacher1988} provides a very powerful technique to determine dissociating autoionizing states. This technique enables us to transform the adiabatic PECs obtained from standard electronic structure calculations into diabatic PECs, unraveling the complicated pattern of interactions between states and allowing for the identification of the dissociating states of interest. One can perform conventional  electronic structure calculations of the desired size and accuracy and then, with modest additional effort, obtain diabatic PECs of comparable reliability. (Calculating the PECs of the ion involved in DR is straightforward and requires only standard techniques of quantum chemistry.)

Application of the block diagonalization method requires extra care when determining the
molecular orbitals (MOs) to insure that the variation of the MOs from one geometry to the
next is small and predictable. Standard electronic structure calculations do not routinely
provide such MOs,  so one must carefully craft the MOs for each geometry.  Section \ref{sec:gettingMOs} describes how we determined suitable MOs.

\subsection{Electronic Configurations of SH$^+$ and SH} \label{sec:SHconfigs}

The ground electronic state of \SHplus\ is a $^3\Sigma^-$ state ($^3A_2$ in the $C_{2v}$ point group of our calculations). Using a simplified notation, the electronic configuration can be written as follows:
\begin{equation}
 1s^2 2s^2 2p^6 3s^2 (\mathrm{SH} )^2 \pi_x^1 \pi_y^1
\end{equation}
where (SH) represents an $sp$-hybridized bonding MO. When an electron recombines with \SHplus, an excited doublet or quartet state of SH can be formed. In the present study we examine the SH doublet states.

The doublet states considered include the SH Rydberg state with the
electronic configuration
\begin{equation}
 1s^2 2s^2 2p^6 3s^2 (\mathrm{SH} )^2 \pi_x^1 \pi_y^1 4s^1
\end{equation}
which is of $^2\Sigma^-$ ($^2A_2$) symmetry. It correlates
with the asymptotic limit $\mathrm{S}(3s^2 3p^3 4s \, ^3S^0) + \mathrm{H}(1s)$.

Other doublet states are the Rydberg states of electronic configuration
\begin{equation}
 1s^2 2s^2 2p^6 3s^2 (\mathrm{SH} )^2 \pi_x^1 \pi_y^1 4p^1
\end{equation}
which can be of $^2\Sigma^-$ ($^2A_2)$ or $^2\Pi$ ($^2B_1$ or $^2B_2$) symmetry and correlate to the asymptotic limit $\mathrm{S}(3s^2 3p^3 4p \, ^3P) + \mathrm{H}(1s))$.

The recombination of SH$^+$ with an electron can also lead to the valence states of SH. The lowest one is a $^2\Pi$ state ($^2B_1$ or $^2B_2$)  with electronic configuration
\begin{equation}
 1s^2 2s^2 2p^6 3s^2 (\mathrm{SH})^2 \pi^3
\label{eq:SHGS}
\end{equation}
that correlates with the lowest $\mathrm{S}(3s^2 3p^4 \, ^3P) + \mathrm{H}(1s))$ limit. Correlating with this same limit is the lowest $^2\Sigma^-$  ($^2A_2)$ state, whose electronic configuration is
\begin{equation}
 1s^2 2s^2 2p^6 3s^2 (\mathrm{SH})^2 \pi^2  (\mathrm{SH^*})^1
\end{equation}
and corresponds to the occupation of the SH antibonding orbital (denoted $\mathrm{SH^*})$. This state is repulsive and is the lowest dissociating state of SH. Other electronic configurations have the general form
\begin{equation}
 1s^2 2s^2 2p^6 3s^2 (\mathrm{SH} \,\, \pi  \,\, \mathrm{SH}^*)^5
\end{equation}
where $(\mathrm{SH} \,\, \pi  \,\, \mathrm{SH}^*)^5$ means that five electrons are distributed into the three specified orbitals. These configurations lead either to the next higher excited valence states of SH [the $A \, ^2\Sigma^+$ ($A_1$), $1\, ^2\Delta$ ($A_1$ or $A_2$) and $2 \, ^2\Pi$ ($B_1$ or $B_2$)] and correlate with the $\mathrm{S}(3s^2 3p^4  \, ^1D) +  \mathrm{H}(1s)$ limit or to the $2 \, ^2\Sigma^+$ ($A_1$) state and correlate with the $\mathrm{S}(3s^2 3p^4  \, ^1S) +  \mathrm{H}(1s)$ limit.

\subsection{Determination of Molecular Orbitals} \label{sec:gettingMOs}

\begin{figure*}[t]
\centering
\includegraphics[width=0.7\textwidth]{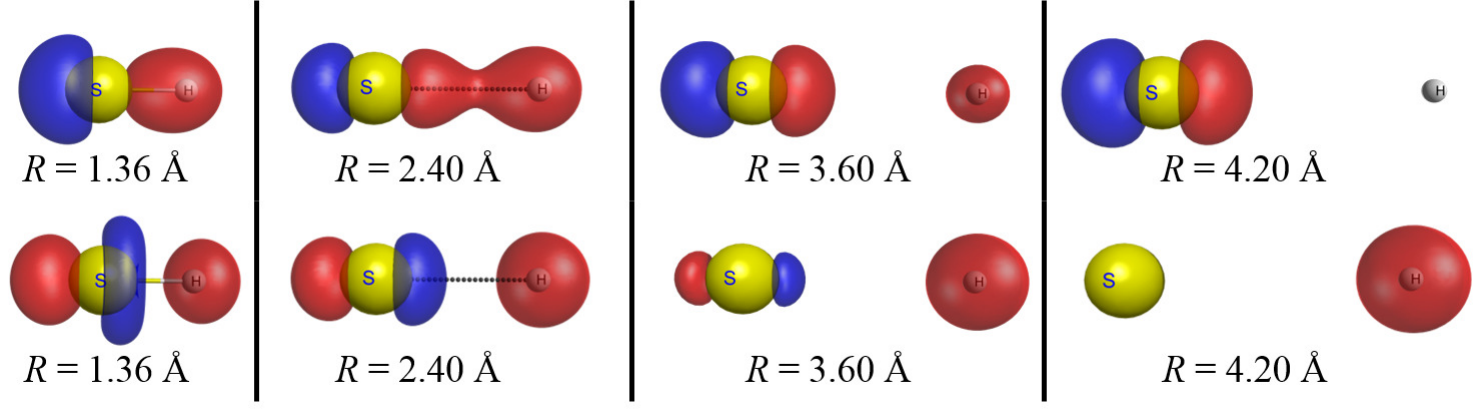}
\caption{The first row of this figure shows four isosurface plots, for different values of $R$, of one of the $A_1$ optimized MCSCF orbitals in the active space. For small $R$ this orbital corresponds to the SH-bonding (SH) orbital. As $R$ increases this orbital smoothly changes to a $3p_z$ atomic orbital centered on S\@. The second row shows isosurface plots for the corresponding antibonding orbital (SH$^{\ast}$).  For large $R$ this orbital becomes a $1s$ atomic orbital centered on H. The two MOs shown above display the greatest variation between the molecular and the separated atom limits. For these MOs, the forms shown at 4.2\,\AA\ have nearly reached the large $R$ limit.}
\label{fig:MOs8-10}
\end{figure*}

We determined smoothly varying MOs corresponding to the orbitals needed to describe the electronic configurations discussed in section \ref{sec:SHconfigs} by performing Multi-Configuration Self Consistent Field (MCSCF) calculations at each geometry of the ground state  neutral SH molecule [Eq.~(\ref{eq:SHGS})]. The MCSCF calculations had five orbitals in the frozen core and nine in the active space. The irreducible representations of these MOs in the $C_{2v}$ point group are given below:
\begin{align*}
\text{Frozen core:~~}   &  \text{3-}A_1, \text{1-}B_1,  \text{1-}B_2                  \\
\text{Active space:~~}  &  \text{5-}A_1, \text{1-}A_2,  \text{2-}B_1, \text{2-}B_2          \\
\end{align*}
The frozen core includes the $n=1$ and 2 orbitals of S, and the active space includes S($3s$), (SH), $\pi_x$, $\pi_y$, (SH$^*$), and the five components of the S($3d$) orbital.
The resulting MCSCF calculations had 3460 configuration state functions (CSFs).

By examining isosurface plots \cite{Bod98} we found that the optimized MCSCF orbitals calculated at $R = 6.00$\,\AA\ corresponded very well with the orbitals identified chemically in section \ref{sec:SHconfigs}. Then, starting at $R = 6.00$\,\AA, we stepped down to smaller values of $R$ by performing each new \mbox{MCSCF} using the orbitals from the previous step as the initial guess. We verified that this procedure gave smoothly varying orbitals by directly calculating pseudo-overlap matrix elements $\mathcal{O}_{ij}$ between the MOs at each geometry and  those at the previously calculated geometry. $\mathcal{O}_{ij}$ is defined as the overlap between the $i^\mathrm{th}$ MO at a given geometry and a ``translated'' orbital formed by using the coefficients of the $j^\mathrm{th}$ MO at the previous geometry. The smoothness of the variation of the MOs is related to the extent to which $\mathcal{O}_{ij} \approx \delta_{ij}$ (the Kronecker delta function). For the calculations we did, we found that $\left|\mathcal{O}_{ii}\right| \ge 0.97$. (We use the absolute value sign because of the occasional unwanted sign changes discussed in section \ref{sec:GeneralStrategy}.)  Also, we found $\left|\mathcal{O}_{i \ne j}\right| \le  0.03$, with typical values $\left|\mathcal{O}_{i \ne j}\right| \lesssim 10^{-4}$. We took these results to be a confirmation that the MOs were sufficiently smooth.

Figure~\ref{fig:MOs8-10} illustrates the smooth variation of the molecular orbitals (SH) and (SH$^*$).  In the chemical region (small $R$), these orbitals are bonding and antibonding, respectively.  As $R$ increases  they make a smooth transition to a $3p_{z}$ centered on S and
a $1s$ centered on H.

\subsection{SOCI Calculations for SH$^+$ ($^3\Sigma^-$) and
SH ($^2\Pi, \, ^2\Sigma^-, \, ^2\Sigma^+$ and $^2\Delta$)} \label{sec:SOCIcalcs}

For a general overview of the behavior of the SH PECs along the dissociation pathway, we performed Multi-Reference Configuration Interaction calculations with Single and Double excitations from the active space (MRCI-SD, hereafter denoted SOCI) on the ground states of SH and \SHplus\ and also on the SH valence excited states listed in section \ref{sec:SHconfigs}, including the $n=4$  lowest Rydberg state. For accurate energy calculations we used the Dunning-type correlation-consistent basis sets \cite{Dunning1989,Woon1993} augmented by $n=4$ diffuse functions on S and an $n=2$ diffuse function on H, that is, the aug-cc-PvTZ basis set as implemented in the GAMESS code \cite{GAMESS} used for the present calculations. This basis set led to $n=4$ molecular Rydberg orbitals with a localized character for all values of $R$, allowing these orbitals to be easily identified all along the dissociation path. Since our atomic basis did not include $3d$ diffuse coefficients, Rydberg $3d$ states were not  calculated. For an equivalent treatment of both the valence and Rydberg states of SH we used an active space of ten orbitals for our SOCI calculations: S($3s$), (SH), $\pi_x$, $\pi_y$, (SH$^*$), S($4s$), S($4p_z$), S($4p_y$), S($4p_x$), and H($2s$). The $1s$, $2s$, and $2p$ orbitals of S were frozen in the CI calculations. The number of CSFs for these calculations was \mbox{$3\,114\,864$}--\mbox{$3\,130\,904$} for SH, depending on the state symmetry, and \mbox{$1\,112\,351$} for the ion.

\begin{figure}[t]
\centering
\includegraphics[width=0.95\columnwidth]{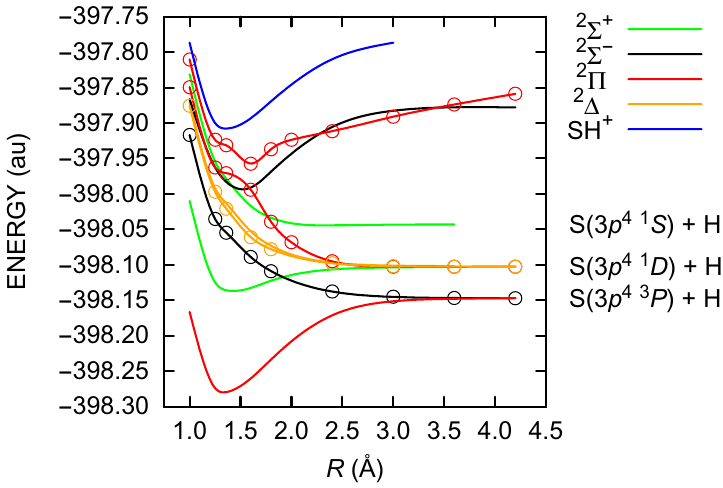}
\caption{Adiabatic SOCI calculations for several symmetries of the SH molecule. These curves were reported previously at a conference \cite{Kas15} and are reprinted with kind permission of The European Physics Journal (EPJ).}
\label{fig:DR2013}
\end{figure}

Figure~\ref{fig:DR2013}  shows the adiabatic SOCI potential energy curves for \SHplus\ ($^3\Sigma^-$) and lowest SH ($^2\Pi$, $^2\Sigma^-$, $^2\Sigma^+$, and $^2\Delta$) states. From our calculations the energy minimum of the ion is located at $R = 1.360\,$\AA, in excellent agreement with the value 1.361\,\AA\ calculated by Khadri \etal \cite{Khadri2006} (and references therein) and the experimental  value  1.364\,\AA\ determined by Dunlavey \etal \cite{Dun79}

The lowest SH state also has its minimum close to this geometry (1.350\,\AA\ from our calculations) in excellent agreement with the calculated values of Bruna and Hirsch \cite{Bruna1987} (1.350\,\AA) and Park and Sun \cite{park1992} (1.347\,\AA) and the experimental value of Dunlavey \etal \cite{Dun79} (1.345\,\AA). Our SH ($^2\Pi$)--$\SHplus(^3\Sigma^-)$ vertical excitation energy  at 1.360\,\AA\ is 10.11\,eV in good agreement with the experimental ionization potential of 10.36\,eV determined by Dunlavey \etal\  \cite{Dun79}. Our $A\,^2\Sigma^+$ state is 3.90\,eV above the ground $^2\Pi$ state, in good agreement with the value $T_e = 3.85$~eV measured by Ramsay \cite{Ramsay1952}. Finally Table~\ref{table:one} shows that our excitation energies involving higher states compare satisfyingly with previously calculated ones. Analysis of our CI wavefunctions confirms the valence nature of the $A\,^2\Sigma^+$ while the second $^2\Sigma^+$ excited state, which is repulsive, has a Rydberg character at short internuclear distances changing to a valence state for geometries above the ion minimum geometry in agreement with the findings of Park and Sun \cite{park1992}. Our second $^2\Sigma^-$ state is a Rydberg state in agreement with previous theoretical assignments \cite{Bruna1987,park1992}. We calculated its excitation energy (adiabatic) from the ground $X\,^2\Pi$ state to be 7.83\,eV to be compared to the 7.37\,eV value reported by Park and Sun \cite{park1992}.

\newcommand{\spacer}{\hspace{2em}}

\begin{table*}
\caption{\label{table:one} Adiabatic vertical excitation energies for the lowest valence states of SH.}
\begin{ruledtabular}
\begin{tabular}{ldddl}
\multicolumn{1}{l}{States} & \multicolumn{1}{c}{This work}                 &  \multicolumn{1}{c}{Bruna and Hirsch  }     &  \multicolumn{1}{c}{Park and Sun}  & \multicolumn{1}{c}{Dissociation limit} \\
                           & \multicolumn{1}{c}{SOCI energies (eV)}        & \multicolumn{1}{c}{(eV)}                     & \multicolumn{1}{c}{(eV)}     \\
                           & \multicolumn{1}{c}{$R=1.360\,\mbox{\AA}$  }  & \multicolumn{1}{c}{$R=1.350\,\mbox{\AA}$  }    \\
\hline
 $X\, ^2\Pi_i$                 & 0.00            &                      &                &  S($3s^23p^4\,\,^3P$) + H \\
 $A \, ^2\Sigma^+$             & 3.90            & 4.05                 & 4.23           & S($3s^23p^4\,\,^1D$) + H \\
 $1\, ^2\Sigma^-$              & 6.11           & 6.19                 &                & S($3s^23p^4\,\,^3P$) + H \\
 $1^2\Delta$: $^2A_1$, $^2A_2$ & 7.24, 7.04     & 7.05, 7.07           &                & S($3s^23p^4\,\,^1D$) +  H \\
 $2 \, ^2\Sigma^-$             & 7.83\footnotemark[1]& 7.57            & 7.37\footnotemark[1]  & S($3s^2 3p^3 4s \,\, ^3S^0$)
 + H \\
 $2 \, ^2\Sigma^+$             & 8.19           & 7.93                 &                & S($3s^23p^4\,\,^1S$) + H \\
 $2 \, ^2\Pi$                  & 8.40           & 8.10                 &                & S($3s^23p^4\,\,^1D$) + H \\
 $3 \, ^2\Pi$                  & 9.47           & 9.37                 &                & S($3s^2 3p^3 4p \,\, ^3P $) + H \\
\end{tabular}
\end{ruledtabular}
\footnotetext[1]{adiabatic excitation energy}
\end{table*}

The feature of the PECs shown in Fig.~\ref{fig:DR2013} most likely to provide a mechanism for DR is the strong avoided crossing between the second and third $^2\Pi$ states of SH. Analysis of the CI wave functions indicates that the $2\,^2\Pi$ state has  Rydberg character for short internuclear distances and valence dissociative character starting from the ion minimum geometry, while the $3 \,^2\Pi$ state has Rydberg character. This strong avoided crossing is the signature for a curve crossing between a repulsive and a bound state not very far from the equilibrium geometry of the ion. Existence of this avoided crossing suggests that the $^2\Pi$ states are prime candidates for curves leading to DR.

Among the other states, the $^2\Sigma^+$ states could conceivably play a role.  However, the lowest  $^2\Sigma^+$ PEC is rather smooth and seems unlikely to contribute significantly to DR through an avoided crossing with a higher $^2\Sigma^+$ state.  Also, the observed DR branching ratio to the S($^1S) +$~H asymptotic limit (0.6\%) is very small~\cite{Bec16thesis}.  Since the $^2\Sigma^+$ states correlate with this asymptotic limit, one can infer that states of this symmetry are unlikely to be important for DR.  The $^4\Pi$ states are more promising candidates.  Bruna and Hirsch \cite{Bruna1987} found a strong avoided crossing between the first two states of that symmetry.

After considering the $^2\Pi$ and $^4\Pi$ states, we decided to focus on the $^2\Pi$ states of SH for our initial investigation.  The ground electronic state of SH is $^2\Pi$, and ample data are available~\cite{NIST_ASD} for validating our potential curve calculations before eventually extending our approach to the quartets.

\subsection{FOCI Calculations}  \label{sec:FOCIcalcs}

As we analyzed the SOCI calculations it became apparent that it would not be practical to use this method to determine the desired dissociating autoionizing states.  The size of the active space, the number of CSFs, and the number of roots required would lead to extremely large computational requirements. Therefore we decided to switch from SOCI (MRCI - singles and doubles) to FOCI (MRCI - singles only) and to enlarge the active space by adding the five spherical components of the $3d$ polarization MO [optimized together with the S($3s$), (SH), $\pi_x$, $\pi_y$ (SH$^*$) MOs through the same MCSCF procedure described above]. This procedure was found to be effective in our previous study \cite{Kas15} of the DR of $\mathrm{N_2H^+}$ with electrons; the very large active space compensates for the lower order CI.

The final FOCI calculations had five frozen core, 15 active space, 48 virtual, and five frozen virtual MOs.  The irreducible representations of these MOs are given below for the $C_{2v}$ point group:
\begin{align*}
\text{Frozen core:~~}   &  \text{3-}A_1, \text{1-}B_1,  \text{1-}B_2                  \\
\text{Active space:~~}  &  \text{8-}A_1, \text{1-}A_2,  \text{3-}B_1, \text{3-}B_2    \\
\text{Virtual space:~~} & \text{18-}A_1, \text{6-}A_2, \text{12-}B_1, \text{12-}B_2
\end{align*}
The five highest-energy virtual space MOs (3-$A_1$, 1-$B_1$, and 1-$B_2$)
were frozen out of the calculation. The resulting calculation had $2\,020\,158$ CSFs and up to 50 roots were needed to identify the autoionizing states of interest in the region of the ion minimum.

We verified the accuracy of the above procedure by calculating the asymptotic limits of the $^2\Pi$ excited states, showing that our results were close to the values reported in the NIST database~\cite{NIST_ASD} (see Table~\ref{table:two}). For example, at the FOCI level we calculated that the S($^1D$) and the Rydberg S($^3P$) atomic states (which correspond to the asymptotic $2\,^2\Pi$ and $3\,^2\Pi$ states of SH) lie respectively $1.226\,$eV and $8.077\,$eV above the
ground state atomic S($^3P$) that is correlated to the asymptotic SH $X\,^2\Pi$. These numbers agree extremely well with the corresponding excitation energies reported in the NIST database~\cite{NIST_ASD} ($1.145$~eV and $8.045$~eV, respectively.) This agreement is very satisfying since the three lowest $^2\Pi$ states of SH are the states important for the present study. Discrepancies are observed for the higher atomic states with differences of 0.15--0.26~eV, which is in the accepted range for high atomic excited states especially in the case of S \cite{Bac08}.

\begin{table*}
\caption{\label{table:two} FOCI Excitation energies at the asymptotic limits for the
SH($^2\Pi$) state compared to the sulfur atomic excitation energies from
the NIST database.}
\begin{ruledtabular}
\begin{tabular}{lddd}
\multicolumn{1}{c}{Electronic state} &
     \multicolumn{1}{r}{FOCI (eV)}  &
           \multicolumn{1}{r}{NIST (eV)}  &
               \multicolumn{1}{r}{$\Delta$ ($\mathrm{FOCI}-\mathrm{NIST}$) (eV)} \\
  &  \multicolumn{1}{r}{$R=6.00\,$\AA}  \\
\hline
S($3s^2 3p^4\,\,^3P$) + H  \rule{0pt}{10pt}                                  & 0.000000   & 0.000000   & 0.000000  \\
S($3s^2 3p^4\,\,^1D$) + H  \rule{0pt}{10pt}                                  & 1.226495  & 1.145442  & 0.081054  \\
S$\left( 3s^2 3p^3 (^4S^0) 4p   \,\, ^3P  \rule{0pt}{10pt} \right)$ + H      & 8.076918  & 8.045206  & 0.031712  \\
S$\left( 3s^2 3p^3 (^2D^0) 4s   \,\, ^3D  \rule{0pt}{10pt} \right)$ + H     & 8.561869  & 8.408156  & 0.153713  \\
S$\left( 3s^2 3p^3 (^4S^0) 4s   \,\, ^1D  \rule{0pt}{10pt} \right)$ + H        & 8.844611  & 8.584403  & 0.260208  \\
S($3s 3p^5\,\,^3P$) + H  \rule{0pt}{10pt}                                   & 9.057100  & 8.929775  & 0.127325  \\
\end{tabular}
\end{ruledtabular}
\end{table*}

Table~\ref{table:three} shows the generally good agreement between our FOCI and SOCI
calculations at the large internuclear distances. The only disagreement is
for the $\mathrm{S}(3s^2 3p^3 4p \, ^3P) + \mathrm{H}$ limit. The analysis of the
CI wave function of the $3\,^2\Pi$ state at this large internuclear distance
shows a contribution from $3d$ polarization orbitals. These orbitals are not
in the active space of the SOCI calculations because of calculation-size practicality.
The discrepancy suggests that these MOs are important for the description
of the $3\,^2\Pi$ state.

\begin{table}
\caption{\label{table:three} Comparison of $^{2}\Pi$ excitation energies at the asymptotic
limits obtained from the SOCI and the FOCI calculations.}
\begin{ruledtabular}
\begin{tabular}{ld}
\multicolumn{1}{c}{Electronic states}       &  \multicolumn{1}{l}{$\Delta$ (FOCI$-$SOCI) } \\
\multicolumn{1}{c}{$R = 6\,\mbox{\AA}$ } &\multicolumn{1}{l}{\hspace{3em}(eV)}                  \\
\hline
S($3s^2 3p^4\,\,^3P$) + H  \rule{0pt}{10pt}                                         & 0.0000   \\
S($3s^2 3p^4\,\,^1D$) + H  \rule{0pt}{10pt}                                         & 0.0188  \\
S$\left( 3s^2 3p^3 (^4S^0) 4p   \,\, ^3P  \rule{0pt}{10pt} \right)$ + H             & -0.3133   \\
S$\left( 3s^2 3p^3 (^2D^0) 4s   \,\, ^3D  \rule{0pt}{10pt} \right)$ + H             & 0.0979   \\
S$\left( 3s^2 3p^3 (^4S^0) 4s   \,\, ^1D  \rule{0pt}{10pt} \right)$ + H             &-0.0078   \\
S($3s 3p^5\,\,^3P$) + H  \rule{0pt}{10pt}                                                  \\
S$^+$ + H$^-$           \rule{0pt}{10pt}                                           & 0.0947 \\
S$^-$ + H$^+$            \rule{0pt}{10pt}                                           & 0.0226 \\
\end{tabular}
\end{ruledtabular}
\end{table}

At the FOCI level our $X \,^2\Pi$--$2 \, ^2\Pi$ and $X \, ^2\Pi$--$3 \, ^2\Pi$  vertical excitation energies (at 1.360\,\AA) are 8.24\,eV and 9.31\,eV, in excellent agreement with our SOCI values and with the results of Bruna and Hirsch \cite{Bruna1987}  (Table~\ref{table:one}). The results obtained at the both FOCI and SOCI levels are very close and FOCI excitations energies compare extremely well with available data~\cite{Bruna1987,NIST_ASD}, justifying calculating the PECs needed for the diabatization at the FOCI level.

Our FOCI adiabatic potential energies curves for the $^2\Pi$ states of SH together with the PEC of the ion are shown in Fig.~\ref{fig:FOCI_DB} in section \ref{sec:BLOCKDappSH}. The adiabatic curves exhibit multiple avoided crossings. The next section describes the diabatization of these curves in order to determine dissociating autoionizing states.

\section{Determination of Diabatic Potential Curves using Block Diagonalization}
\label{sec:BLOCKD}

\subsection{Formalism}     \label{sec:BLOCKDformalism}

\begin{figure}[t]
\centering
\includegraphics[width=0.9\columnwidth]{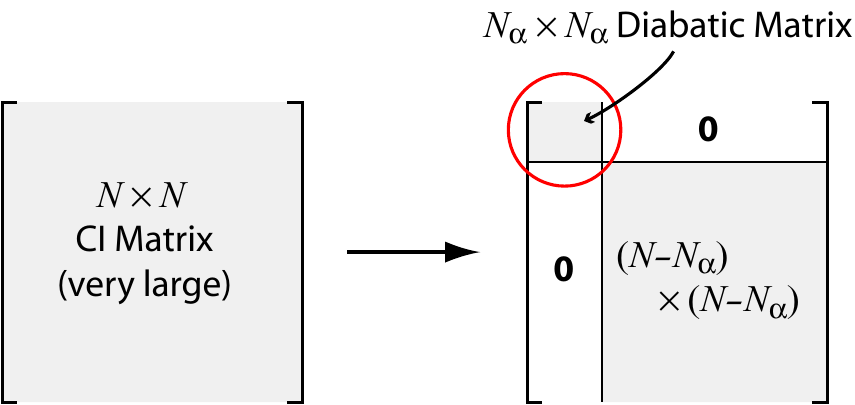}
\caption{\label{fig:BlockDiag} Schematic illustration of block diagonalization. The $N_\alpha \times
N_\alpha$ block is the desired diabatic matrix. The larger block of size $N - N_\alpha$
need not be explicitly determined.}
\end{figure}

This section presents a brief summary of the block diagonalization method \cite{Dom93,Spi00,Hic05,Hic11,Kashinski2012}. We assume that the electronic structure calculations have provided a set of adiabatic electronic states whose wave functions $\Psi_i(R)$ can be expressed as linear combinations of configuration state functions (CSFs) $\Phi_j(R)$ constructed from MOs with a consistent chemical interpretation at all $R$:
\begin{equation}  \label{eq:definecij}
\Psi_i(R) = \sum_{j=1}^N c_{ij}(R)\Phi_j(R)
\end{equation}
Ideally, the largest coefficients $c_{ij}$ for the electronic  states of interest (eigenvectors $\Psi_i$) come from a small set of specific CSFs ($\Phi_j$) that correspond to well-defined electronic configurations. Then one must identify a set of $N_\alpha$ CSFs that make the dominant contribution to $N_\alpha$ electronic states of interest. The total number of CSFs ($N$) is typically very large, possibly of order $10^6$, while $N_\alpha$ is much smaller, perhaps 5--25. Then one can determine a diabatic matrix of size $N_\alpha \times N_\alpha$ by transforming the original Hamiltonian matrix determined by the configuration interaction (CI) calculation to block diagonal form.  The transformation is illustrated schematically in Fig.~\ref{fig:BlockDiag}. The diagonal elements of the $N_\alpha \times N_\alpha$ matrix correspond to the diabatic potential energies, and the off-diagonal elements to the coupling between  the diabatic electronic states.

The first step of the procedure is to construct an $N_\alpha \times N_\alpha$ matrix $\mathbf{S}$ by selecting the coefficients $c_{ij}$ [defined in Eq.~(\ref{eq:definecij})] for the $N_\alpha$ dominant configurations in the $N_\alpha$ states of interest. Each column of $\mathbf{S}$ contains the $N_\alpha$ appropriate CSF coefficients for one of the $N_\alpha$ adiabatic electronic states. Then the diabatic Hamiltonian matrix \Hdia\ can be expressed as a transformation of the diagonal matrix $\mathbf{E}$ whose nonzero elements are the adiabatic eigenvalues $E_1, \ldots, E_{N_\alpha}$:
\begin{equation}\label{eq:HisTET}
     \Hdia = \mathbf{T}^\dag \mathbf{E} \mathbf{T},
\end{equation}
where ($^\dag$) denotes the adjoint (transpose for a real transformation), and
\begin{equation} \label{eq:defineT}
     \mathbf{T} = \mathbf{S}^{-1} \left( \mathbf{S} \mathbf{S}^\dag
     \right)^{1/2}.
\end{equation}

Block diagonalization has two very desirable features.  Since the transformation defined by Eqs.~(\ref{eq:HisTET}) and (\ref{eq:defineT}) is unitary, the eigenvalues of the diabatic matrix are exactly the original adiabatic energies.  Also, an important numerical consideration is that the diabatic matrix can be determined using only operations on matrices of size $N_\alpha \times N_\alpha$. (Of course, the desired energies and eigenvectors of the large CI matrix must be computed by standard electronic structure techniques, which do involve larger matrices.)

\subsection{Application to SH}  \label{sec:BLOCKDappSH}

\subsubsection{Connection of \Hdia\ to DR}

In the present case the CSFs that define the diabatic Hamiltonian \Hdia\ can be identified either with Rydberg states, which are described by attractive potential curves roughly parallel to the \SHplus\ ion curve, or with valence states, which tend to be dissociating curves.  Both types of curves were illustrated in Fig.~\ref{fig:DRexplain2}.  Since the dissociating curve can intersect many Rydberg states, the interactions between the diabatic states can lead to a complicated pattern of avoided crossings of adiabatic states.  In the diabatic representation, the off-diagonal elements of \Hdia\ provide the Rydberg-valence coupling.  Using the ideas of quantum defect theory \cite{seaton1983}, one can scale this coupling to obtain an estimate of the coupling \Vel\ between a dissociating, autoionizing state and an electron-molecular-ion scattering state:
\begin{equation}    \label{eq:Vel}
\Vel = (n^*)^{1.5}
     \langle \Psi_\mathrm{Rydberg} | \Hel  | \Psi_\mathrm{dissoc}\rangle,
\end{equation}
where $n^* = n - \mu$ is the effective quantum number of the Rydberg state determined by its binding energy relative to the parent ion; $\mu$ is the quantum defect, and \Hel\ is the electronic Hamiltonian. This scaling  will be used  in section \ref{subsec:disspec} to estimate the coupling terms needed for DR from \Hdia.

\subsubsection{General strategy} \label{sec:GeneralStrategy}

The first step in the diabatization is to select the $N_\alpha$ CSFs that define the diabatic states.  This process requires careful consideration and judgment.  Which CSFs have the largest coefficients often depends on the internuclear separation, causing uncertainty about how to achieve a consistent treatment for all geometries.  Also, it often seemed that the number of CSFs that contributed strongly to a given set of eigenvectors was larger than the number of
eigenvectors considered.

A technique that proved very useful was to define certain linear combinations of CSFs as ``super CSFs''. This procedure, which simply amounts to a change of basis in a linear vector space, often led to a much clearer
interpretation of the wavefunctions.  The super CSFs were easy to identify because certain small sets of CSFs corresponding to the same orbital occupancies often appeared with the same relative coefficients in several different eigenvectors and at several values of $R$.  In the simplest case, two CSFs, one can relate the relative coefficients to a mixing angle $\theta$ and then define
\begin{align}
\Phi^+ &= \phantom{-}\cos\theta\,\Phi_{1} + \sin\theta\,\Phi_{2} \notag \\
\Phi^- &= -\sin\theta\,\Phi_{1} + \cos\theta\,\Phi_{2}
\end{align}
We found other cases where as many as five CSFs with the same orbital occupancies appeared repeatedly with very similar coefficients. Then we determined a more general orthogonal transformation.  The super CSFs usually corresponded to the correct electronic spin configuration for the state of interest. Defining these super CSFs enabled us to reduce the total number of CSFs (or super CSFs) needed to define the diabatic states.

Another important consideration was which adiabatic electronic states to include in the diabatization.  At small values of $R$, where the dissociating state is very repulsive and can mix with many Rydberg states, a very large number of adiabatic states can have significant contributions from the diabatic CSFs. In order to provide quantitative guidance for the selection of the adiabatic states, we developed a computer code that calculated the magnitude of the projection of each adiabatic state in the space spanned by the diabatic CSFs.  By trying several sets of CSFs with different values of $N_\alpha$, and selecting the best adiabatic states for each set, we could systematically select appropriate parameters for the diabatization.

One of the uncertainties in the calculation was caused by the dependence of the coefficients of some of the CSFs on the internuclear separation $R$. As $R$ changed, some of the high-energy eigenvectors became more and less important and were therefore swapped in and out of the diabatization.  Whenever one of the higher adiabatic states in the diabatization changed, there was an unavoidable discontinuity in the diabatic curves. However, since these higher adiabatic states were only indirectly coupled to the dissociating states of interest, the effect was slight, and we did extensive testing with different numbers of states to eliminate any anomalies.

Another uncertainty in the diabatization was caused by the difficulty in assigning the sign of the off-diagonal matrix elements of \Hdia.  At several stages of the calculation, GAMESS checks the normalization of the MOs and sets the largest coefficient of each MO to be positive.  Since the relative magnitude of the MO coefficients may depend on the internuclear separation, abrupt sign changes occasionally appeared as a function of $R$.  We were usually able to eliminate these unwanted sign changes by inspection, unless the matrix elements were very small.

\subsubsection{Results for \Hdia\ with  ground-core Rydberg state (R$_1$)}\label{sec:HdiaC1}

\begin{figure*}
\centering
\includegraphics[width=0.85\textwidth]{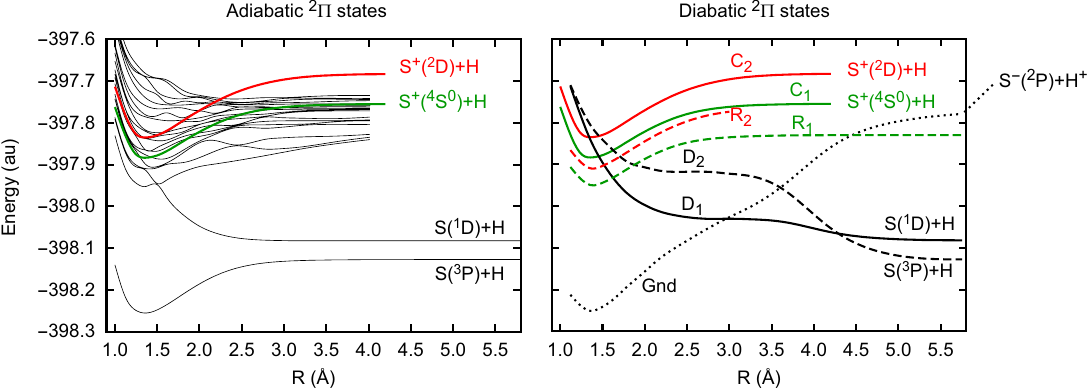}     
\caption{Comparison between adiabatic and diabatic curves for SH.  The left panel shows the splined adiabatic FOCI potential curves for the $^2\Pi$ states of the SH molecule, and the right panel shows the corresponding diagonal diabatic curves for SH. Both panels show the two lowest adiabatic curves for SH$^+$, the ground state ($^3\Sigma^-$)  in green and the first excited state ($^1\Delta$) in red. The black states in the left panel exhibit many avoided crossings typical of adiabatic curves. The nature of the various interacting states is much more clear after the transformation to diabatic states: the right panel shows two dissociating states (D$_1$ and D$_2$), two ion core states (C$_1$ and C$_2$), and the lowest Rydberg state for each core (R$_1$ and R$_2$).  D$_1$,  C$_1$ and R$_1$ correspond to the states shown schematically in Fig.~\ref{fig:DRexplain2}. The curve labelled Gnd corresponds to the ground adiabatic state at small $R$ and an ion pair state at large $R$.  This state does not play a role in the electron capture that leads to DR.}
\label{fig:FOCI_DB}
\end{figure*}

Two types of Rydberg states were considered in this work: one with the ground state ($^3\Sigma^-$) ion core and one with the first excited state ($^1\Delta$) ion core. These curves merit special attention because of the importance of the Rydberg-valence coupling for the DR process.  This section focuses on the first case.  In the region of small $R$ near the minimum of the \SHplus\ potential, the lowest SH($^2\Pi$) Rydberg states are built on the ground state ion core and have the following electronic configuration, where the vertical line ``$|$'' denotes the end of the ion core orbitals:
\begin{equation}
  1s^2 2s^2 2p^6 3s^2 (\mathrm{SH})^2 \pi_x^1 \pi_y^1 \, |\, (\mathrm{SH}^*)^0 4p^1
\end{equation}
Here the bonding orbital (SH) is doubly occupied; the antibonding orbital (SH$^*$) is empty, and performing the diabatization is very straightforward.  By inspection of the eigenvectors in this range we identified a set of 16 CSFs (or super CSFs) and 16 adiabatic states that led to sensible diabatic states.  This same model also led to good diabatic, dissociating curves at all values of $R$.  However, additional effort was needed to determine the behavior of the Rydberg diabatic potential curves for large $R$.

As $R$ increases,  the bonding and antibonding SH orbitals change as shown in Fig.~\ref{fig:MOs8-10}. In the separated atom limit,  (SH) becomes $3p_z$ on S and is singly occupied, and (SH$^*$) becomes $1s$ on H, also singly occupied.  Two distinct CSFs are required to represent accurately both the molecular region and the separated atom limit.  To obtain accurate diabatic potentials for all values of $R$, we included an additional super CSF in the diabatization to represent the asymptotic limit of the lowest Rydberg state correctly.  This procedure, however, leads to two diabatic curves, one correct in the molecular limit at small $R$, and one correct at large $R$.  Therefore at each $R$ we diagonalized the $2 \times 2$ submatrix of the originally-calculated \Hdia.  This produced the final Rydberg potential curve with the correct behavior at all values of $R$.

Another issue arose in the calculation of the off-diagonal elements of \Hdia\ involving the lowest Rydberg state. Some of these matrix elements did not vanish in the limit of large $R$.  The reason can be understood by considering Table~\ref{table:three}, which gives the electronic configuration of the lowest few $^2\Pi$ states of SH at large $R$. The lowest two asymptotes are the dissociating states $D_2$ and $D_1$, which correlate with H($1s$) plus S($^3P$) and S($^1D$), respectively.  The next state is the lowest Rydberg $R_1$ and correlates with H($1s$) plus a Rydberg S($^3P$) state.  In the diabatization, we represent each of the states at large $R$ by a single CSF.  In this case the matrix element of \Hdia\ between the two dissociating states will be zero because of the different symmetries of the S atom.  Similarly, the matrix element of \Hdia\ between the CSFs corresponding to $D_1$ state and the $R_1$ state must be zero.  However, the matrix element between $D_2$ and $R_1$ need not be zero, because the diabatization uses a single CSF for each of those states.  In practice, we found that the matrix element was small, but not zero, so we adopted the following procedure: At every $R$, we applied an additional, constant (independent of $R$) $2 \times 2$  orthogonal transformation to the diabatic matrix, chosen to diagonalize exactly the submatrix of \Hdia\ corresponding to the $D_2$ and $R_1$ diabatic states.   \Hdia\ then had the correct asymptotic behavior.

\begin{figure}[t]
\centering
\includegraphics[width=0.95\columnwidth]{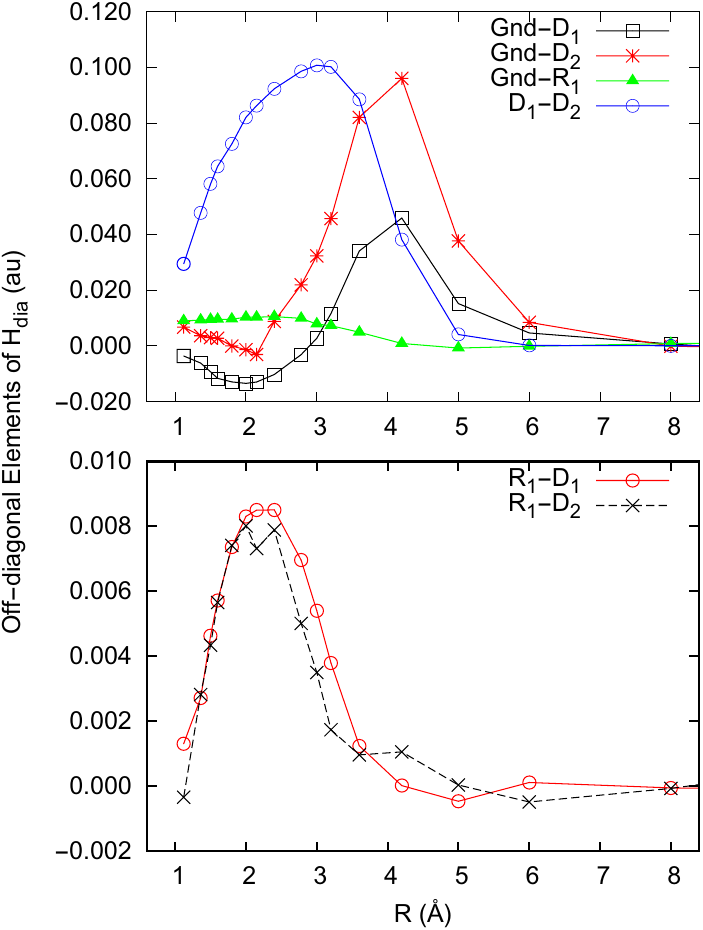}
\caption{Off-diagonal elements of \Hdia. The matrix elements in the lower panel involve the coupling between the dissociating states and the Rydberg states and play an important role in DR. Note the difference in scale between the top and bottom panels.}
\label{fig:OffDiagA}
\end{figure}

Figure \ref{fig:FOCI_DB} summarizes our results for the diagonal diabatic states and for the adiabatic states whose calculation was described in section \ref{sec:FOCIcalcs}.  The block diagonalization transforms the strongly interacting adiabatic states shown in the left panel to the smoothly-varying and easily-identifiable diabatic states in the right panel.  The diabatic states in this figure are the diagonal elements of \Hdia; the coupling between the diabatic states  labelled Gnd, $R_1$, $D_1$ and $D_2$ correspond to the off-diagonal elements of \Hdia\ and are shown in Fig.~\ref{fig:OffDiagA}.  Coupling terms involving R$_2$ will be discussed in the next section.

\subsubsection{Results for \Hdia\ with a core-excited Rydberg state (R$_2$)}

\begin{figure}
\centering
\includegraphics[width=0.95\columnwidth]{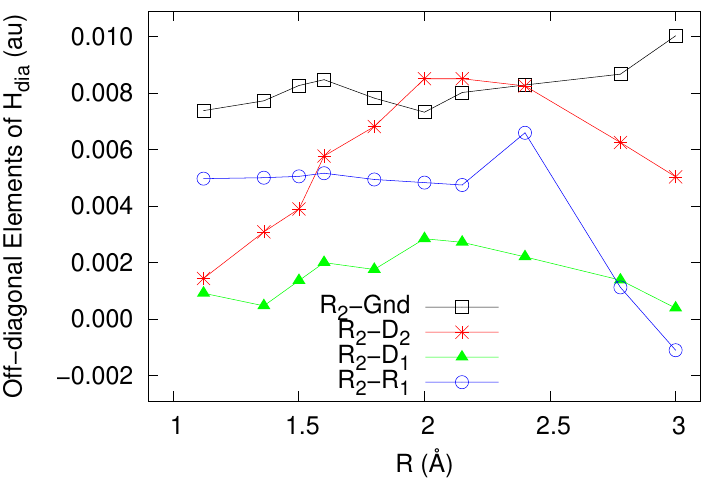}   
\caption{Off-diagonal elements of \Hdia\ involving the core-excited Rydberg state.}
\label{fig:OffDiagB}
\end{figure}

The excited $\SHplus (1 \, ^1\Delta)$ state can support Rydberg states, and we also considered these core-excited Rydberg states.  In the molecular region the lowest of these states is a linear combination of the electronic configurations
\begin{equation}
  1s^2 2s^2 2p^6 3s^2 (\mathrm{SH})^2 \pi_x^2  \pi_y^0 \, |\,(\mathrm{SH}^*)^0 4s^1
\end{equation}
and
\begin{equation}
  1s^2 2s^2 2p^6 3s^2 (\mathrm{SH})^2 \pi_x^0  \pi_y^2 \,|\, (\mathrm{SH}^*)^0 4s^1.
\end{equation}
As with the lowest Rydberg levels, adding core-excited Rydberg levels to the diabatization was straightforward in the region of small $R$.  Treating the separated-atom limit was also straightforward; in this case the electronic configuration was a $1s$ H atom plus S in the following configuration:
\begin{equation}
1s^2 2s^2 2p^6 3s^2 3p_x^1 3p_y^1  3p_z^1  \,|\,4s^1
\end{equation}
We were able to describe the asymptotic wave function in a satisfactory manner with one super CSF, but obtaining reliable results for the off-diagonal elements of \Hdia\ at intermediate values of $R$ proved to be difficult.  The results were not robust in this region; small changes in the parameters of the diabatization led to significant changes in the off-diagonal curves. For this reason, we only report the matrix elements of \Hdia\ involving the core-excited Rydberg state for values of $R \le 3.0$~\AA.  The diagonal potential for this state was shown in Fig.~\ref{fig:FOCI_DB}, and off-diagonal elements are given in Fig.~\ref{fig:OffDiagB}.

\section{The  MQDT-type approach to DR}  \label{sec:mqdt}

The MQDT approach~\cite{seaton1983,gj85,jungen96,giusti80} has been shown to be a powerful method for the evaluation of the cross sections of the DR and competitive processes like ro-vibrational and dissociative excitations. It was applied with great success to several diatomic systems like H$_2^+$ and its isotopologues \cite{giusti83,ifs-a09,takagi93,tanabe95,ifs-a18,amitay99,epee2015}, O$_2^+$ \cite{ggs91,guberman-dr99}, NO$^+$ \cite{sn90,ifs-a22,ifs-a36,Ngassam1,motapon06b}, CO$^+$~\cite{mjzs2015}, N$_2^+$~\cite{jtx}, BF$^+$~\cite{mjzs2016} and triatomics like H$_{3}^{+}$ \cite{ifs-a26,kokoou01,kokoou03}.

\subsection{The theoretical approach to the dynamics}  \label{subsec:grcore}

The theoretical summary given below is limited to an account of the {\it vibrational} structures and couplings for the ion cores, illustrated mainly for DR~(Eq.~(\ref{eq:reac-dr})). However, the reader should keep in mind that the other competitive reactions---such as {\it superelastic collision} (SEC) ($v^+_i>v^+_f$, where $v_{i}^{+}$ and $v_{f}^{+}$ stand for the initial and final vibrational quantum number of the target ion), {\it elastic collision} (EC) ($v^+_i=v^+_f$) and {\it inelastic collision} (IC) ($v^+_i<v^+_f$)---occur simultaneously and display quite similar features.

The DR results from the quantum interference between the \textit{direct} mechanism involving the autoionizing  resonant states SH$^{**}$  and the \textit{indirect} one occurring via Rydberg predissociating states  SH$^{*}$.

A detailed description of our theoretical approach was given in \cite{mjzs2015,jtx}. Its main steps are the following:

\textit{i) Building of the interaction matrix} $\boldsymbol{\mathcal{V}}$:  For a given symmetry $\Lambda$ of the neutral system, and assuming that one single partial wave of the incident electron contributes to the relevant interactions, the $R-$dependent electronic coupling of an ionization
channel relying on the electronic-core state $c_\beta$ ($\beta = 1$ for the ground state and  $\beta = 2$ for the first-excited state) with the dissociation channel $d_j$ can be written:
\begin{equation}\label{eq:elcoup}
\MV^{(e)\Lambda}_{d_{j},c_\beta}(R) =
\langle\Phi_{d_j}|H_{el}|\Phi^{el,c_\beta}\rangle, \beta = 1, 2.
\end{equation}
where $\MV^{(e)\Lambda}_{d_{j},c_\beta}(R)$ is assumed to be independent of the energy of the external electron, and the integration is performed over the electronic coordinates of the neutral (electron plus ion({\it core})) system. Here \Hel\ denotes the electronic Hamiltonian; $\Phi_{d_j}$ is the electronic wave function of the dissociative state, and $\Phi^{el,c_\beta}$ is the wave function describing the molecular system ``Rydberg electron + ion in its $c_\beta$ electronic state''.

Similarly, the electronic coupling between the two ionization continua is:
\begin{equation}\label{eq:elcoupCores}
\MV^{(e)\Lambda}_{c_1,c_2}(R) =
\langle\Phi^{el,c_1}|H_{el}|\Phi^{el,c_2}\rangle.
\end{equation}

Integrating these couplings over the internuclear distance leads to the non-vanishing elements of the interaction matrix $\V(E)$:
\begin{align}
V_{d_j,v_{c_\beta}}^{\Lambda}(E) & =  \langle F_{d_j}(E)|\MV^{(e)\Lambda}_{d_{j},c_\beta}(R)|\chi_{v_{c_\beta}}
	\rangle,   \quad \beta = 1, 2.     \label{eq:Vdv}
 \\
 & V_{v_{c_1},v_{c_2}}^{\Lambda}  = \langle \chi_{v_{c_1}}|\MV^{(e)\Lambda}_{c_1,c_2}(R)|\chi_{v_{c_2}}
	\rangle.     \label{eq:Vvv}
\end{align}
Here $\chi_{v_{c_\beta}}$ ($\beta$ = 1, 2) is the vibrational wave function associated with an ionization channel relying on the core $c_\beta$; $F_{d_j}$ is the regular radial wave function of the dissociative state $d_j$, and $E$ is the total energy of the molecular system. This interaction is effective at short electron-ion and internuclear distances typical of the reaction zone.

\textit{ii) Computation of the reaction matrix} $\boldsymbol{\mathcal{K}}$, by adopting the second-order perturbative solution of the Lippman-Schwinger integral equation~\cite{Ngassam2,florescu2003,motapon06b}, written in operator form as
\begin{equation}\label{eq:solveK}
\boldsymbol{\mathcal{K}}= \boldsymbol{\mathcal{V}} + \boldsymbol{\mathcal{V}}{\frac{1}{E-\boldsymbol{H_0}}}\boldsymbol{\mathcal{V}},
\end{equation}
where $\boldsymbol{H_0}$ is the Hamiltonian of the molecular system under study, with the inter-channel interactions neglected.

The reaction matrix $\Kmat$ in block form is
\begin{equation}\label{Kmat2}
\Kmat = \left( \begin{array}{ccc}
\Kmat_{\bar d\bar d} & \Kmat_{{\bar d}{\bar v}_{c_1}} & \Kmat_{\bar d {\bar v}_{c_2}}\\
\Kmat_{{\bar v}_{c_1} \bar d} & \Kmat_{{\bar v}_{c_1} {\bar v}_{c_1}} & \Kmat_{{\bar v}_{c_1} {\bar v}_{c_2}}\\
\Kmat_{{\bar v}_{c_2} \bar d} & \Kmat_{{\bar v}_{c_2} {\bar v}_{c_1}} & \Kmat_{{\bar v}_{c_2} {\bar v}_{c_2}}\\
\end{array} \right),
\end{equation}
where the collective indices $\bar d$, $\bar v_{c_1}$, $\bar v_{c_2}$,  span the ensembles of all individual indices $d_j$, $v_{c_1}$ and $v_{c_2}$, which respectively label dissociation channels, ionization channels built on core 1 and ionization channels built on core 2.

An extensive and rigorous derivation of the structure of each block of the $\Kmat$-matrix in second order for a multi-core case was provided in our earlier work~\cite{kc2013a}. For SH$^+$, with two attractive ion cores, a natural application of our earlier work leads to the following form of the $\Kmat$-matrix in second order:
\begin{equation}\label{Kmat2}
\Kmat = \left( \begin{array}{ccc}
\Omat & \V_{\bar d \bar v_{c_1}} & \V_{\bar d \bar v_{c_2}} \\
\V_{\bar v_{c_1} \bar d} & \Kmat_{\bar v_{c_1} \bar v_{c_1}}^{(2)} & \V_{\bar v_{c_1} \bar v_{c_2}}\\
\V_{\bar v_{c_2} \bar d} & \V_{\bar v_{c_2} \bar v_{c_1}} & \Kmat_{\bar v_{c_2} \bar v_{c_2}}^{(2)}\\
\end{array} \right).
\end{equation}
where the elements of the diagonal blocks of $\Kmat$ are

\begin{widetext}         
\begin{equation}
K_{v_{c_\beta} v'_{c_\beta}}^{{\Lambda}(2)} =   \sum_{d_j} \frac{1}{W_{d_j}}
\int \int \Big[\chi_{v_{c_\beta}}^\Lambda(R)\MV^{(e)\Lambda}_{c_\beta,d_{j}}(R)F_{d_j} (R_<) G_{d_j}(R_>)
\MV^{(e)\Lambda}_{d_{j},c_\beta'}(R')
\chi_{v'_{c_\beta}}^\Lambda
(R') \Big] \, dR \, dR',  \qquad
\beta = 1,2
\end{equation}
and where $W_{d_j}$ is the Wronskian of the pair ($F_{d_j}$, $G_{d_j}$), the latter being the irregular internuclear wave function associated with the dissociative curve $d_j$ at the given total energy of the system.

\textit{iii) Computation of the eigenchannel wavefunctions} based on the eigenvectors  and eigenvalues of the reaction matrix $\boldsymbol{\mathcal{K}}$, i.e. the columns of the matrix $\boldsymbol{{U}}$ and the elements of the diagonal matrix $\boldsymbol{\tan(\eta)}$ respectively :
\begin{equation}\label{eq:EqnForU}
\boldsymbol{\mathcal{K}U}= -\frac{1}{\pi}\boldsymbol{\tan(\eta)U},
\end{equation}
where the non-vanishing elements of the diagonal matrix $\boldsymbol\eta$ are the phaseshifts introduced into the wavefunctions by the short-range interactions.

\textit{iv) Frame transformation from the Born-Oppenheimer representation to the close-coupling one} is performed via the matrices $\boldsymbol{\mathcal{C}}$ and $\boldsymbol{\mathcal{S}}$, built on the basis of the matrices $\boldsymbol{{U}}$ and $\boldsymbol\eta$ and on the  quantum defect characterizing the incident/Rydberg electron, $\mu_{l}^{\Lambda}(R)$. The elements of these matrices are
\begin{align}
\mathcal{C}_{v^+_{c_\beta},\Lambda \alpha} & =
     \sum_{v_{c_\beta}} U^{\Lambda}_{{v_{c_\beta}}, \alpha}
     \left\langle \chi_{v^+_{c_\beta}} (R)\left| \cos\left(\pi\mu^{\Lambda}_{c_\beta} (R) + \eta_{\alpha}^{{\Lambda}}\right) \right| \chi_{v_{c_\beta}}(R) \right\rangle,    \qquad \beta=1,2   \label{C1}
\\
\mathcal{C}_{{d_j},\Lambda\alpha} & = U^{\Lambda}_{{d_j}, \alpha} \cos\eta^{\Lambda}_\alpha \label{C2}
\\
\mathcal{S}_{v^+_{c_\beta},\Lambda \alpha}  & =
     \sum_{v_{c_\beta}} U^{\Lambda}_{{v_{c_\beta}}, \alpha}
     \left\langle \chi_{v^+_{c_\beta}} (R)\left| \sin\left(\pi\mu^{\Lambda}_{c_\beta} (R) + \eta_{\alpha}^{{\Lambda}}\right) \right|\chi_{v_{c_\beta}}(R) \right\rangle,   \qquad \beta=1,2  \label{S1}
\\
\mathcal{S}_{{d_j},\Lambda\alpha}  & = U^{\Lambda}_{{d_j}, \alpha}  \sin\eta^{\Lambda}_\alpha,          \label{S2}
\end{align}
where $\alpha$ denotes the eigenchannels built through the diagonalization of the reaction matrix $\Kmat$.
\end{widetext}        

\textit{v) Construction of the generalized scattering matrix $\boldsymbol{\mathcal{X}}$}, eventually split in blocks associated with open and/or closed (o and/or c respectively) channels:
\begin{equation}
\boldsymbol{\mathcal{X}}=\frac{\boldsymbol{\mathcal{C}}+i\boldsymbol{\mathcal{S}}}{\boldsymbol{\mathcal{C}}-i\boldsymbol{\mathcal{S}}}
\qquad
\boldsymbol{\mathcal{X}}= \left(\begin{array}{cc} \boldsymbol{X_{oo}} & \boldsymbol{X_{oc}}\\
                   \boldsymbol{X_{co}} & \boldsymbol{X_{cc}} \end{array} \right).
\end{equation}

\textit{vi) Construction of the physical scattering matrix $\boldsymbol{\mathcal{S}}$}, whose elements link mutually the open channels exclusively, given by
\cite{seaton1983}:
\begin{equation}\label{eq:solve3}
\boldsymbol{S}=\boldsymbol{X_{oo}}-\boldsymbol{X_{oc}}\frac{1}{X_{cc}-\exp(-i2\pi\boldsymbol{ \nu})}\boldsymbol{X_{co}}.
\end{equation}
\noindent
Here the matrix $\exp(-i2\pi\boldsymbol{ \nu})$ is diagonal and relies on the effective quantum numbers $\nu_{v^{+}}$ associated with the vibrational thresholds of the closed channels.

\textit{vii) Computation of the cross-sections:}
Given the target cation in its level $v_i^+$, its impact with an electron of energy $\varepsilon$ results in dissociative recombination according to the formula:
\begin{equation}\label{eqDR}
\sigma _{{\rm diss} \leftarrow v_{i}^{+}}=\sum_\Lambda \frac{\pi}{4\varepsilon}
\rho^\Lambda\sum_{j}\mid S_{d_{j},v_{i}^{+}}\mid^2,
\end{equation}
where $\rho^\Lambda$ stands for the ratio between the multiplicity of the involved electronic state  and that of the target ion.

\subsection{Molecular data}\label{subsec:disspec}

\begin{table}[b]
\caption{\label{table:four} Fitting parameters $D$ and $C_n$ used in the long-range multipole formula $D + C_n/R^n$ for the different molecular states.}
\begin{ruledtabular}
\begin{tabular}{llccdd}
\multicolumn{2}{c}{Molecular}  &\multicolumn{1}{c}{state} &\multicolumn{1}{c}{$n$}    &  \multicolumn{1}{c}{$D$} & \multicolumn{1}{c}{$C_n$} \\
\multicolumn{2}{c}{system} &\multicolumn{1}{c}{} &\multicolumn{1}{c}{} &\multicolumn{1}{c}{(Hartree)} & \multicolumn{1}{c}{(Hartree$\times \text{Bohr}^n$)}                  \\
\hline
SH$^+$\,\,   &  $^3\Sigma^-$  &C$_1$ & 4   & -399.20323   &       -6.981 \rule{0pt}{10pt}  \\ 
SH$^+$\,\,   &  $^1\Delta$    &C$_2$ & 4   & -399.13556   &      -11.542 \\
SH$^{**}$\,\,&  $^2\Pi$       &D$_1$ & 6   & -399.54164   & 109\,963.040 \\
SH$^{**}$\,\,&  $^2\Pi$       &D$_2$ & 6   & -399.58373   &  14\,902.987 \\
SH$^{*}$\,\, &  $^2\Pi$       &R$_1$ & 6   & -399.28808   &     -746.261 \\
SH$^{*}$\,\, &  $^2\Pi$       &R$_2$ & 6   & -399.26826   &  -1\,232.990 \\
\end{tabular}
\end{ruledtabular}
\end{table}

\begin{figure*}
\centering
\includegraphics[width=0.7\textwidth]{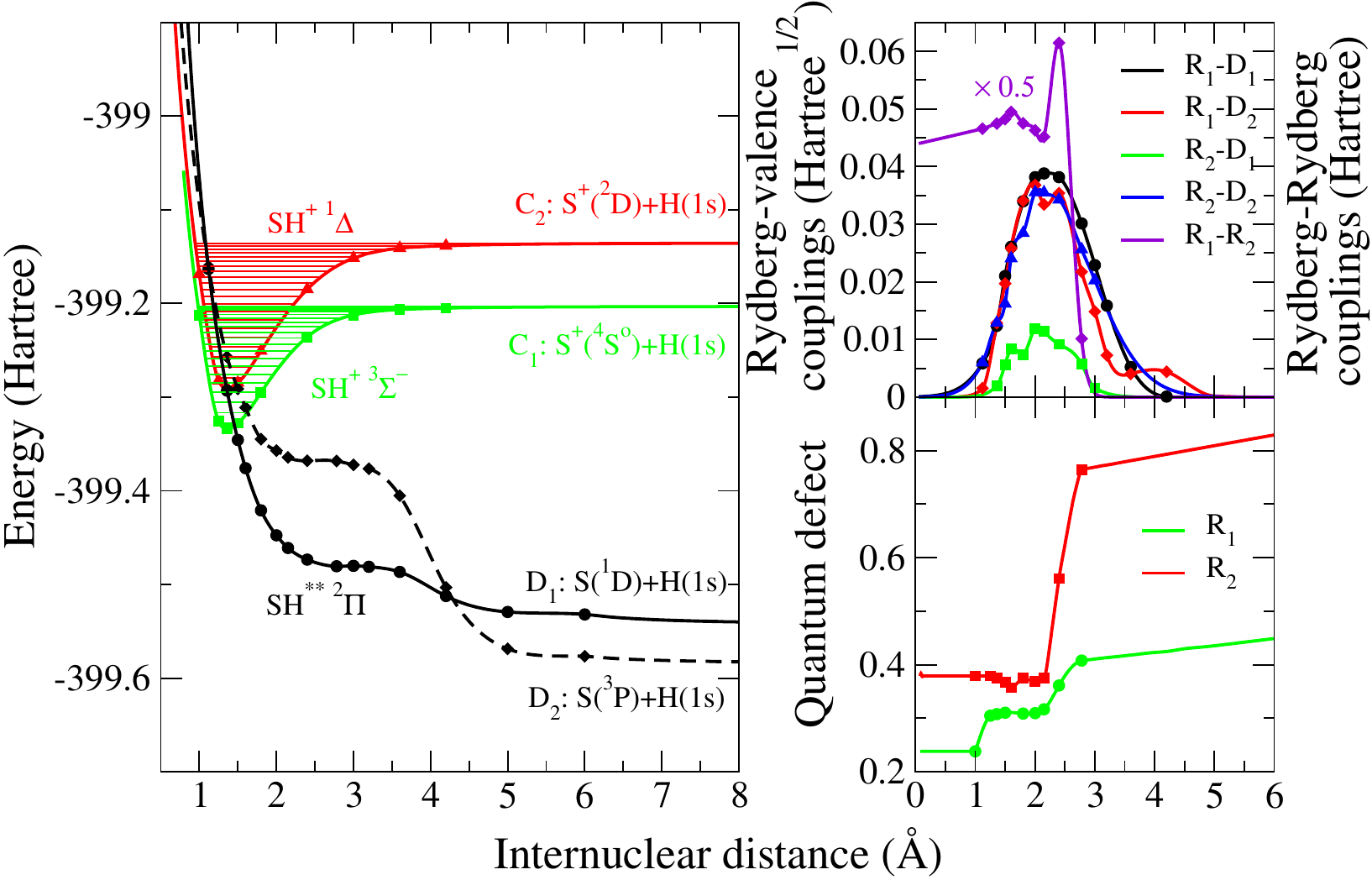}
\caption{Molecular data sets relevant for the dissociative recombination of SH$^+$ compiled from Figs.~\ref{fig:FOCI_DB}--\ref{fig:OffDiagB}. Left panel: the green and red continuous lines stand for the ground (C$_1$: $^3\Sigma^-$) and first excited (C$_2$: $^1\Delta$) electronic states of the ion, having the NIST dissociation limits, whereas the black continuous (D$_1$) and dashed (D$_2$) lines give the dissociative autoionizing states ($^2\Pi$) of the neutral. Upper right panel: electronic coupling: black, red, green and blue between the valence states (D$_1$ and D$_2$) with the Rydberg states (R$_1$ and R$_2$) built on the ground and excited ion cores (C$_1$ and C$_2$) respectively. The violet line gives the couplings between the two Rydberg series. Lower right panel: quantum defects for the Rydberg series based on the ground and excited ion cores. The symbols stand for the quantum chemistry data.}
\label{fig:SHqch}
\end{figure*}

The MQDT treatment of DR and its competitive processes requires data for PECs of the ground and excited ion states, for the relevant dissociative autoionizing states of the neutral molecule, as well as for each series of Rydberg states (in order to determine the quantum defects characterizing these series). Moreover, accurate data on the electronic couplings between the states of the neutral system---bound or continuum with respect to the ionization---are essential.

In addition, the DR cross section is extremely sensitive to the position of the neutral dissociative states with respect to that of the target ion. A slight change of the crossing point between these PECs can lead to a significant change in the predicted DR cross section. In addition, the PECs of the dissociative states must also converge to the correct asymptotic limits for large values of the internuclear distance.

 Several  methods are available to provide all the necessary molecular data with the desired accuracy.  Among these are  R-matrix theory~\cite{tennyson2010}, the complex Kohn variational method~\cite{rescigno1995}, quantum defect methods~\cite{jungen96,gj85, giusti80,giusti83,ifs-a09,ggs91,guberman-dr99,Gub95,florescu2003}, analysis of spectroscopic data~\cite{jungen96,gj85,giusti80,mjzs2015,ifs-a26} and the block diagonalisation method~\cite{pacher1988} presented in more detail in the previous sections.

Figure~\ref{fig:SHqch} summarizes all the molecular data required by MQDT.  The diabatic potential energy curves provided by the block diagonalisation method (D$_1$ and D$_2$ in Fig.~\ref{fig:FOCI_DB}) are given by full symbols in the left panel of Fig.~\ref{fig:SHqch}. The potential curves shown, as well as the Rydberg curves $R_1$ and $R_2$ used to determine the quantum defects, were extended towards large internuclear distances by adding appropriate long range forms $D + C_n/R^n$ defined by the parameters $D$ and $C_n$ given in Table~\ref{table:four}. Moreover, in order to get the NIST atomic dissociation limits we have performed a global shift of $-1.45$ a.u for each of the PECs by preserving all other characteristics (e.g. ionization energies) of the electronic states.

From the PECs of the Rydberg states provided by the block diagonalisation method (R$_1$ and R$_2$ in Fig.~\ref{fig:FOCI_DB}) we have extracted the two sets of smooth quantum defects for the two ion cores, defining the Rydberg series of the $^2\Pi$ symmetry, shown in the lower right panel of Fig.~\ref{fig:SHqch}. 

The upper right panel of Fig.~\ref{fig:SHqch} gives the electronic couplings of the Rydberg states R$_1$ and R$_2$ to the dissociative continuum, as well as the coupling between the two series of Rydberg states. These coupling terms are the $\MV^{(e)\Lambda}_{d_j,c_\beta}(R)$ and $\MV^{(e)\Lambda}_{c_1,c_2}(R)$ defined in Eqs.~(\ref{eq:elcoup}) and  (\ref{eq:elcoupCores})  and are determined using Eq.~\ref{eq:Vel} from the off-diagonal elements of \Hdia\ presented in Figs.~\ref{fig:OffDiagA} and \ref{fig:OffDiagB}. For values of the internuclear distances $R\sim$~2.5--4~\AA\ we fit the coupling terms with gaussian functions chosen to match the peak values of the original results computed by quantum chemistry methods. Since the original results approached zero for large values of $R$ (as discussed in section \ref{sec:HdiaC1}), this procedure provided smooth functions with the correct asymptotic behavior. We have extrapolated the couplings for small internuclear distances in a similar way.

\subsection{Cross sections and rate coefficients}\label{subsec:xs}

In this section we present our results for the cross sections and rate coefficients for the dissociative recombination of vibrationally relaxed SH$^+$ ($v_i^+ = 0$).

Our main objective here is to illustrate the importance of multi-core effects in addition to the existence of multiple dissociative states. Thus we present our results in a progressive way: First we take into account only the ground electronic state of the ion core  and one dissociative valence state of the neutral (C$_1$ and D$_1$). Second, we consider the inclusion in the MQDT calculation of the electronically excited state of the ion core (C$_2$). And third, we add in our treatment the effect of a further dissociative state (D$_2$).

The present calculations include a total of 43 ionization channels associated with 22 vibrational levels of the SH$^+$ X$^{3}\Sigma^{-}$ ground state (C$_1$) and 21 vibrational levels of SH$^+$ a$^{1}\Delta$ excited state (C$_2$). They were performed in the second order of the K-matrix, with the inclusion of both {\it direct} and {\it indirect} mechanisms. For the incident electron energy we explored the range $0.01$ meV $-$ $3.5$ eV and, in terms of electronic temperature, we focused on the range 10--1000~K, relevant for the cold interstellar media.

\begin{figure}
\centering
\includegraphics[width=.95\columnwidth]{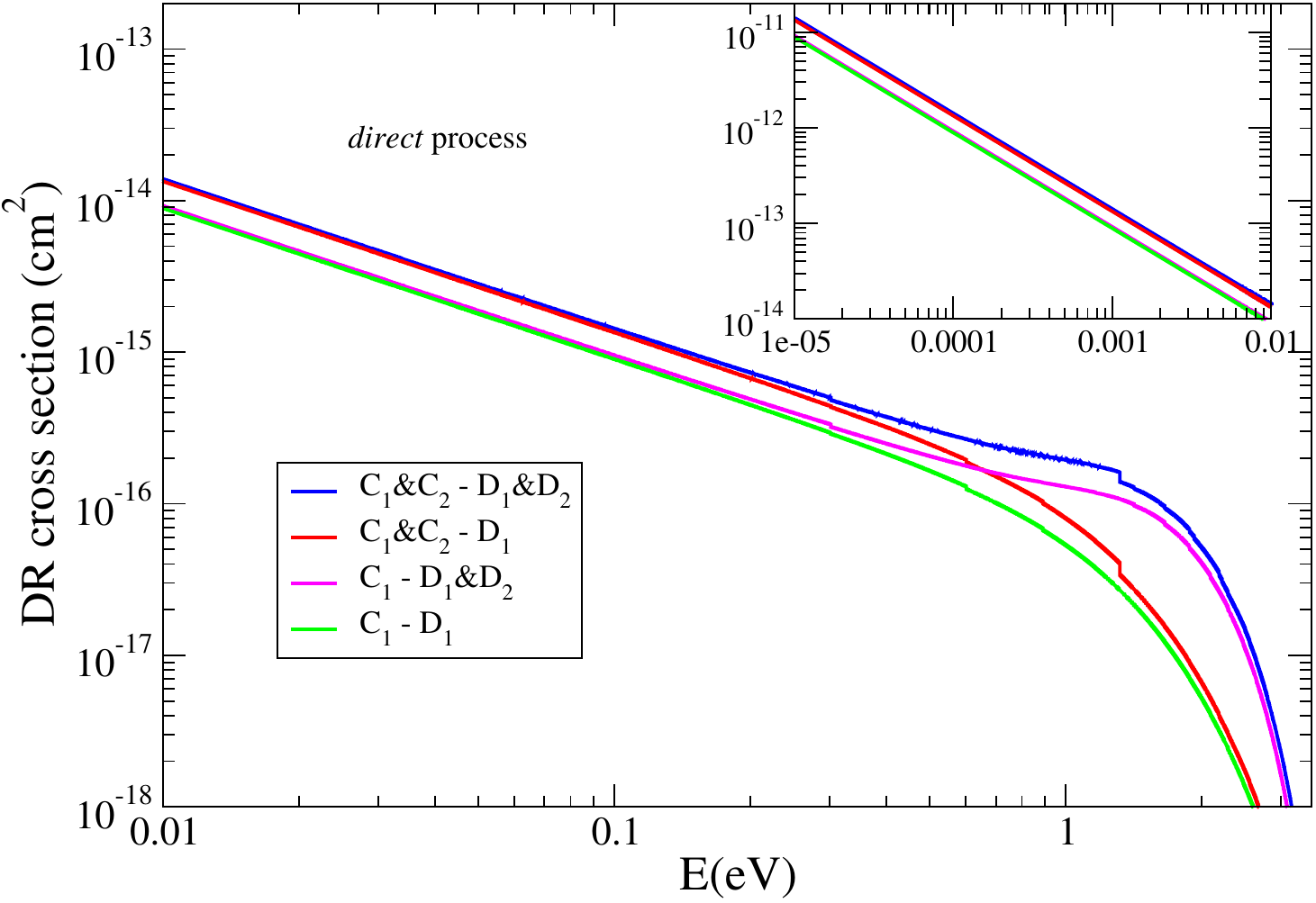}  
\caption{DR cross sections for vibrationally relaxed SH$^+$ for the {\it direct} mechanism only. Significance of colors: green for ground ion core (C$_1$) and one dissociative state (D$_1$); red for ground and excited ion cores (C$_1$ and C$_2$) and one dissociative state (D$_1$); blue for two ion cores (C$_1$ and C$_2$) and two dissociative states (D$_1$ and D$_2$); magenta for ground ion core (C$_1$) and two dissociative states (D$_1$ and D$_2$).}
\label{fig:DRdircs}
\end{figure}

\begin{figure}
\centering
\includegraphics[width=0.9\columnwidth]{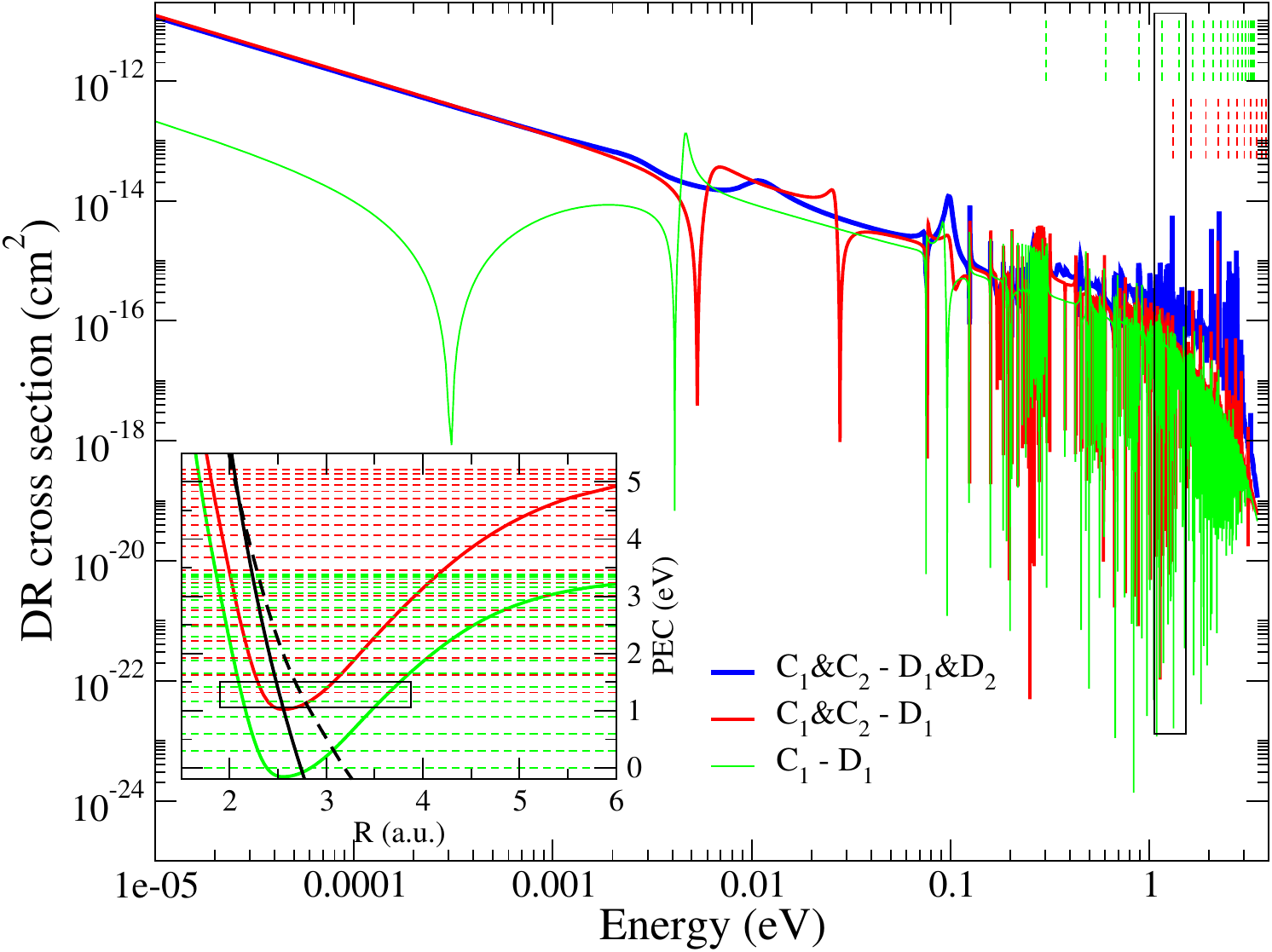}  
\caption{DR cross sections for vibrationally relaxed SH$^+$ for the total ({\it direct} and {\it indirect}) mechanism.  The significance of the colors is the same as in Fig.~\ref{fig:DRdircs}. The inset shows the energy diagram of the system - PECs and vibrational energy levels (vertical and horizontal lines): green for ground ion core (C$_1$), red for excited core (C$_2$), continues black for the D$_1$ dissociative states, dashed black for D$_2$. The rectangular frame indicates the energy range explored in detail in Figs.~\ref{fig:DRcscoreD1} and \ref{fig:DRcscoreD2}.}
\label{fig:DRcs}
\end{figure}

\begin{figure}
\centering
\includegraphics[width=0.95\columnwidth]{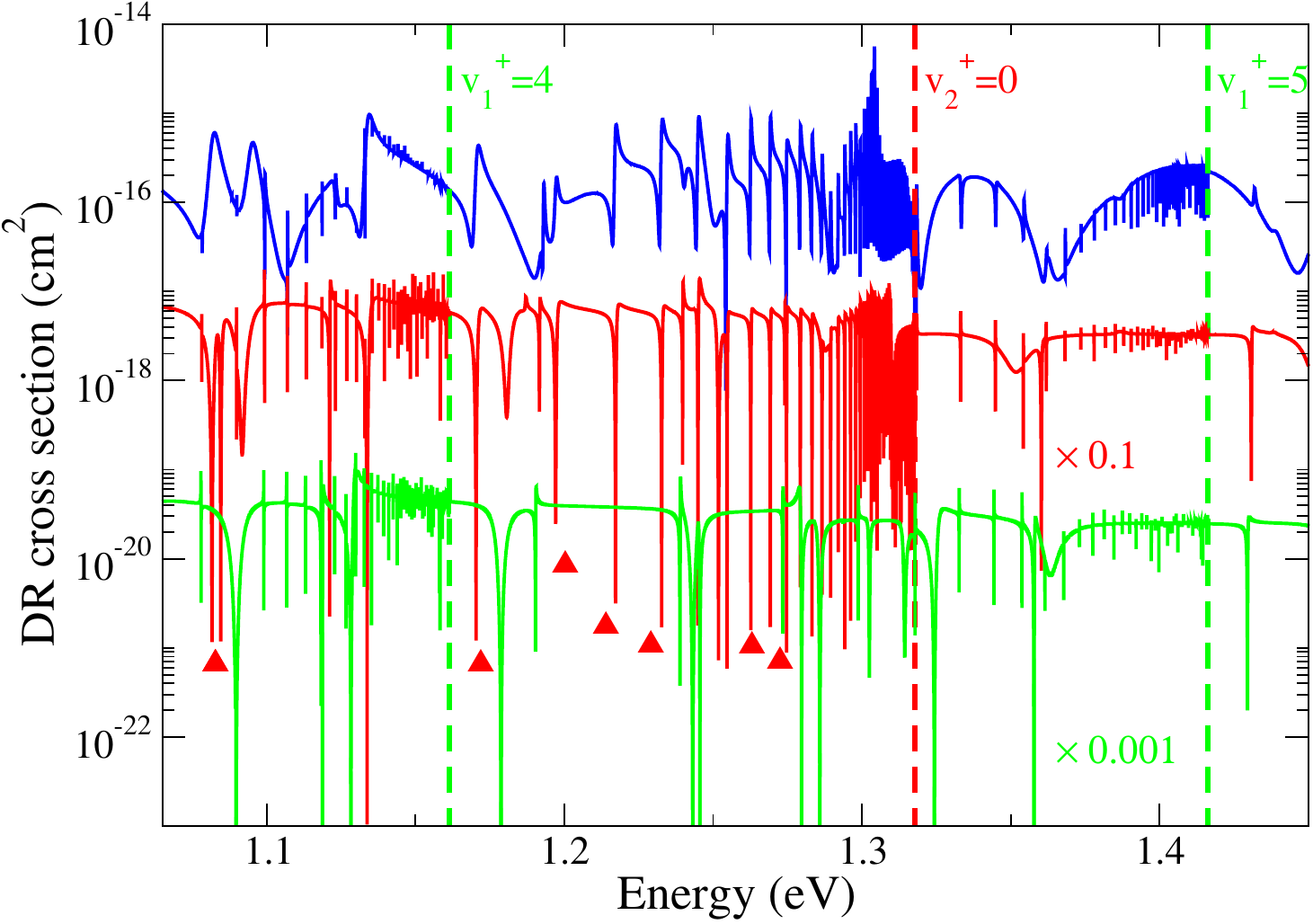}
\caption{Multi-core effects in the DR cross sections for vibrationally relaxed SH$^+$ in the energy range shown in the black rectangle in Fig.~\ref{fig:DRcs}. The significance of the colors is the same as in Fig.~\ref{fig:DRcs}. The cross sections were rescaled with the factors given on the figure in order to have a clearer view of their structures.  The vertical colored lines correspond to the vibrational levels of the ion cores, and the red triangles indicate the resonances associated with the capture into Rydberg states with an excited core.}
\label{fig:DRcscoreD1}
\end{figure}

\begin{figure}
\centering
\includegraphics[width=0.95\columnwidth]{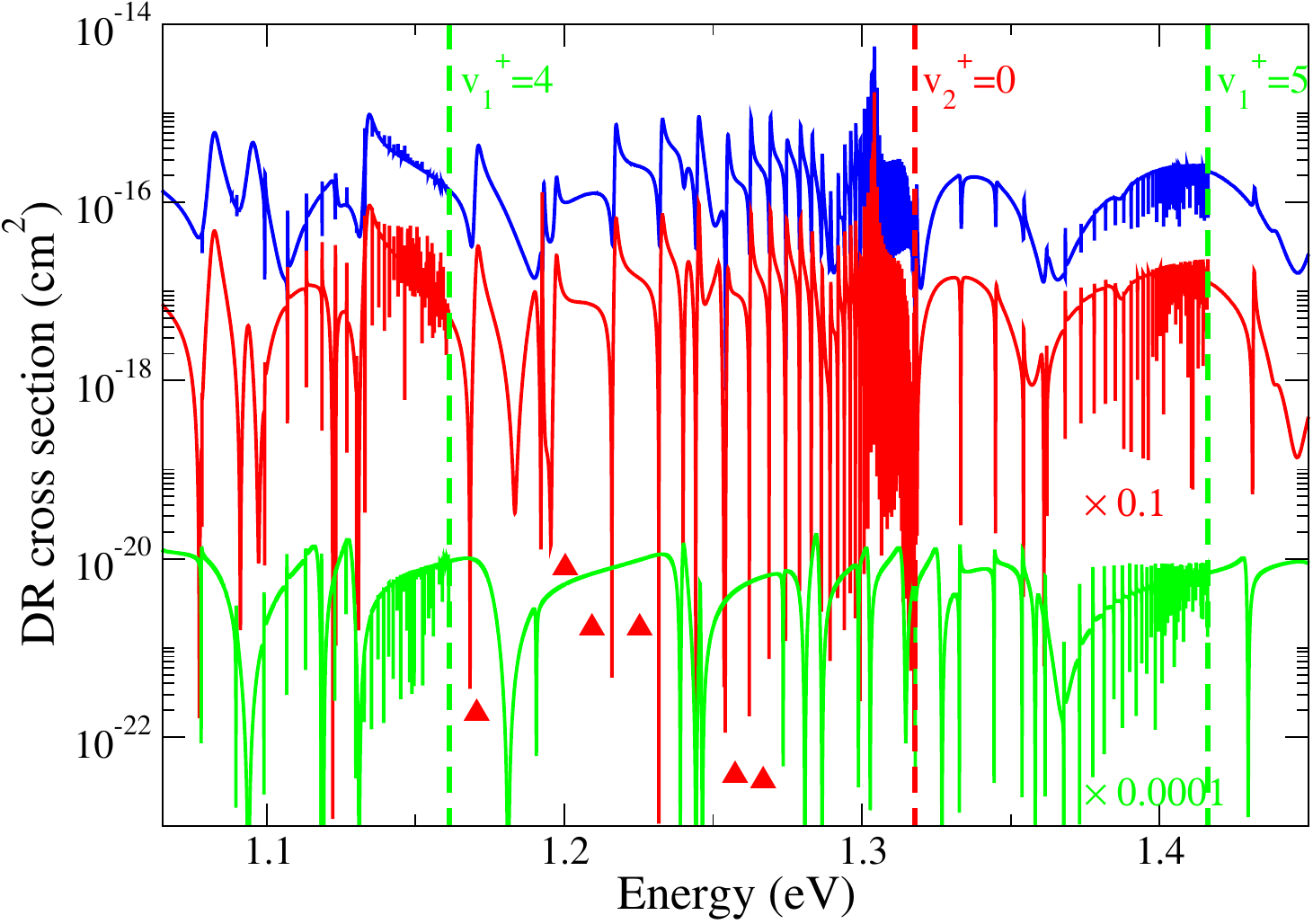}
\caption{Same as Fig.~\ref{fig:DRcscoreD1}, with the dissociative state D$_2$ instead of D$_1$.
}
\label{fig:DRcscoreD2}
\end{figure}

\begin{figure}
\centering
\includegraphics[width=0.9\columnwidth]{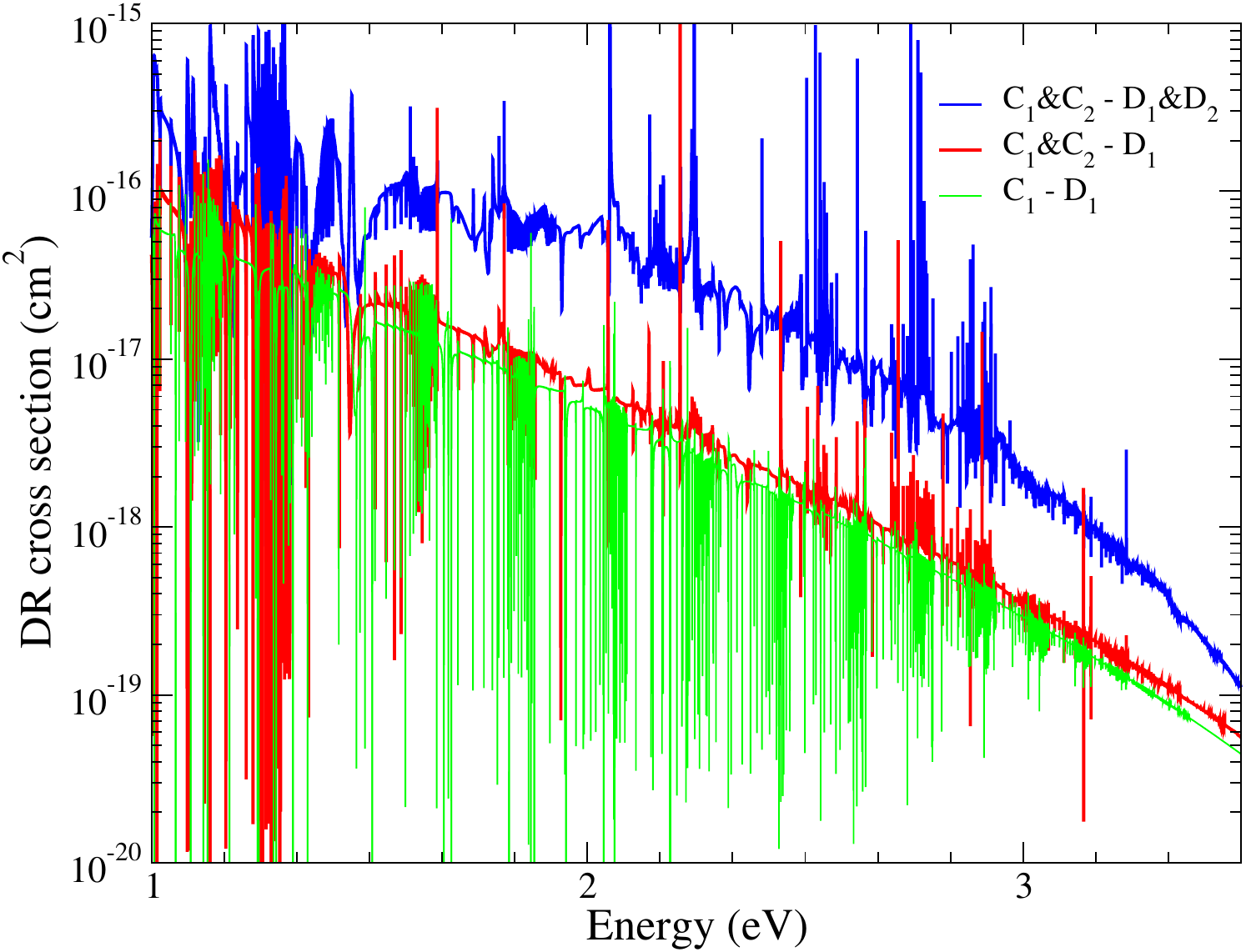}  
\caption{\label{fig:DRcshe} The role of the D$_2$ dissociative state in the DR cross sections for vibrationally relaxed SH$^+$ at high energies. The significance of colors is the same as in Fig.~\ref{fig:DRcs}.}
\end{figure}

\begin{figure}
\centering
\includegraphics[width=0.9\columnwidth]{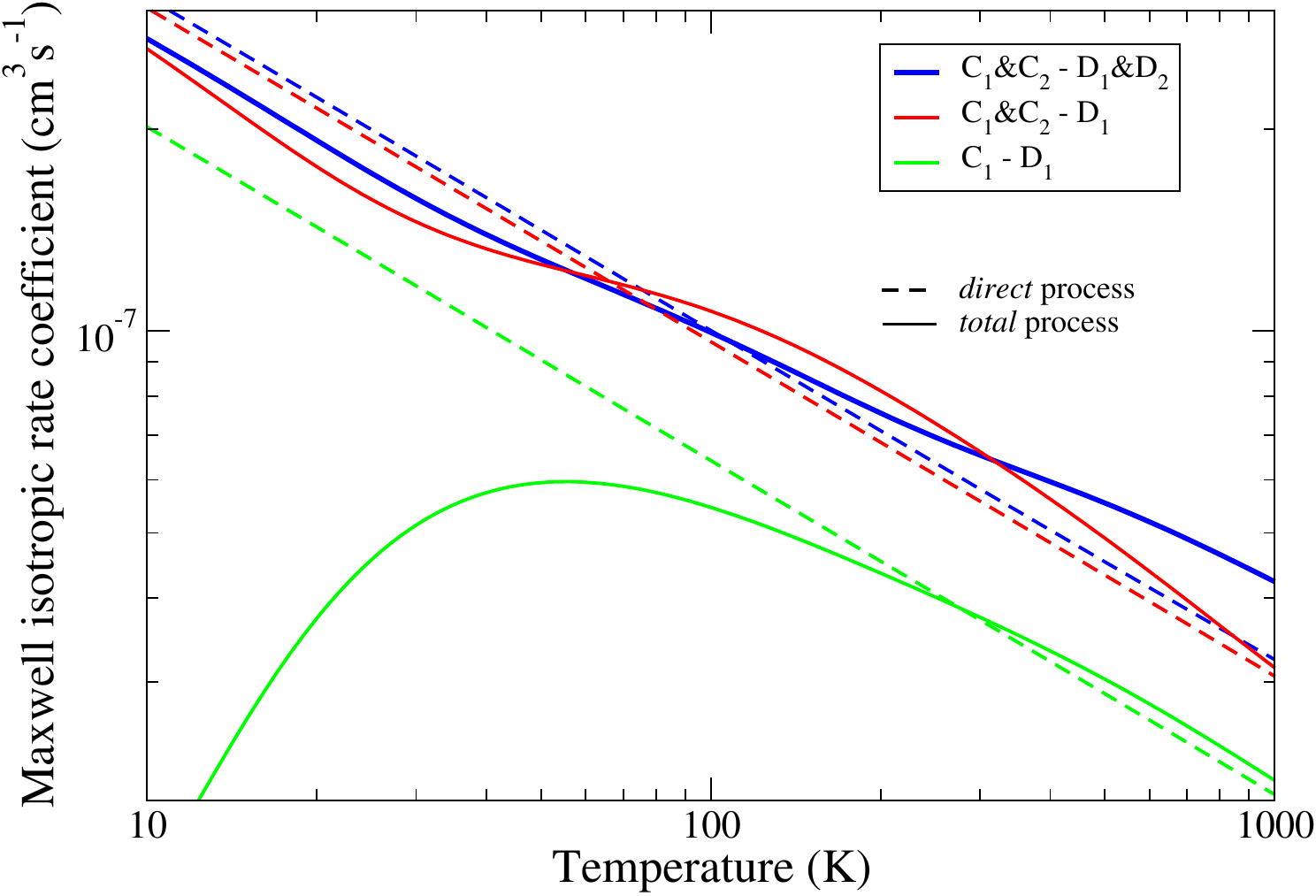}  
\caption{DR Maxwell rate coefficients for vibrationally relaxed SH$^+$.  The dashed lines stand for the {\it direct} process only, while the continuous lines for the {\it total} ones.
}
\label{fig:DRrate}
\end{figure}

\begin{figure}
\centering
\includegraphics[width=0.95\columnwidth]{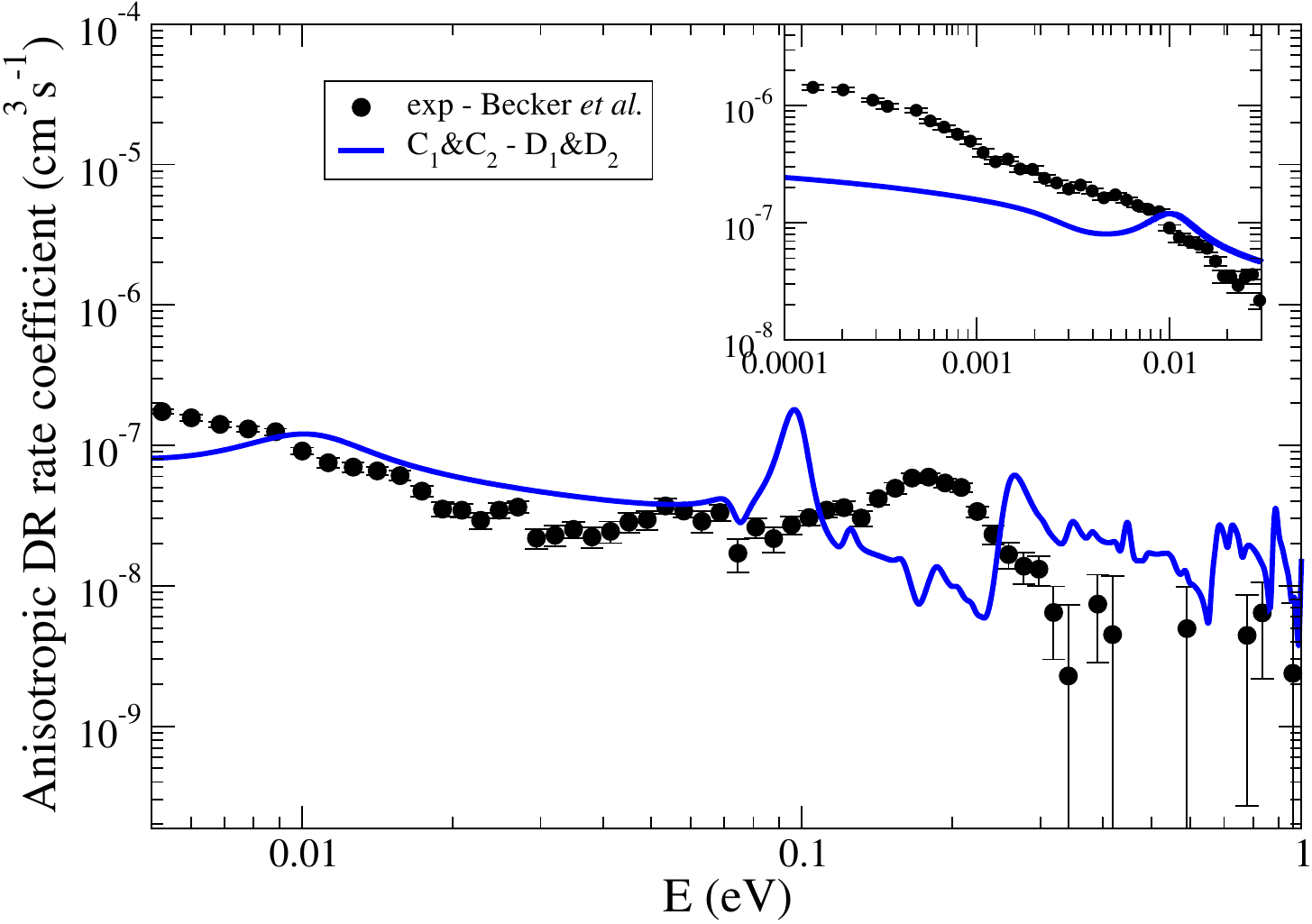}  \caption{ Anisotropic DR rate coefficients for vibrationally relaxed SH$^+$ in the energy range from $0.1$ meV to $1$ eV. The blue solid line stands for the present MQDT results for the ground and excited cores and two dissociative states (D$_1$ and D$_2$).
The full black circles with error bars show the experimental results performed on the TSR storage ring at Heidelberg, Germany~\cite{becker2015}. The longitudinal and transversal temperatures were taken as $kT_\parallel = 25$ $\mu$eV and $kT_\perp = 1.65$ meV. The inset shows the rate coefficients in the low energy range.}
\label{fig:DR2coreanirate}
\end{figure}

Figures~\ref{fig:DRdircs}--\ref{fig:DRcshe} show our calculated DR cross sections, while Fig.~\ref{fig:DRrate} shows the Maxwell DR rate coefficients. Finally, in Fig.~\ref{fig:DR2coreanirate}, our MQDT results are compared with the most recent storage-ring experiments performed at the TSR in Heidelberg~\cite{becker2015}. The color code used is related as follows to the states included in the calculations: green stands for ground ion core (C$_1$) and one dissociative state (either D$_1$ or D$_2$), red  for ground and excited ion cores (C$_1$ and C$_2$) and one dissociative state (either D$_1$ or D$_2$), magenta for ground ion core (C$_1$) and two dissociative states (D$_1$ and D$_2$), and blue for the full calculation with two ion cores (C$_1$ and C$_2$) and two dissociative states (D$_1$ and D$_2$).

The quantitative characterization of the different contributions (multi-core effects vs.\ multiple dissociative states) in the total cross section is a difficult task due to their resonant character. However, in a first step, a simple and reasonably good overall estimation can be provided  by comparing the DR cross sections for the {\it direct} process only, as one can see in Fig.~\ref{fig:DRdircs}. According to this comparison, the multi-core effects due to the favorable crossing between C$_2$ and D$_1$ are of key importance at low collision energy (up to 1 eV),  while the second dissociative state has a crucial contribution at high collision energies due to the favorable crossing between C$_1$ and D$_2$.

Further steps towards a deeper understanding of the role of different mechanisms rely on comparing the effects of the {\it indirect} process (shown in Fig.~\ref{fig:DRcs}) with the calculations in  Fig.~\ref{fig:DRdircs} that include only the {\it direct} process. At first we took into account only the ground ion core  X$^3\Sigma^-$ (C$_1$) and the lowest dissociative state (D$_1$) correlating to the S$(^1D)+$H$(1s)$ atomic limits (see Fig.~\ref{fig:SHqch}).  The resulting cross section is the green solid line, which differs from the corresponding green curve in Fig.~\ref{fig:DRdircs} by the rich resonance structures. With the exception of very low collision energy, where two profound dips are induced by the lowest excited Rydberg states, the resonant structure in Fig.~\ref{fig:DRcs} is superimposed on the smooth background originating from the {\it direct} process only.  Including the first excited ion core of $^1\Delta$ symmetry (shown in red in Fig.~\ref{fig:SHqch}) in the calculation has a remarkable impact on our cross section, as shown by the red curve in Fig.~\ref{fig:DRcs}. The resonance found  at 0.3~meV in the ground-core (C$_1$) case is narrowed and displaced to 5~meV, due to the interaction between the Rydberg states associated with the ground and excited cores. This effect leads to an increase of one order of magnitude of the total DR cross section at low collision energies (red and green curves in Fig.~\ref{fig:DRcs}), while on the whole energy range the overall gain is about a factor of 4.5. Finally, including the second dissociative state, which correlates to the S$(^3P)+$H$(1s)$ atomic limit and is shown as a dashed black line in Fig.~\ref{fig:SHqch}, increases the cross sections even more, especially in the high energy range, as shown by the blue curve in Fig.~\ref{fig:DRcs}. In comparison with the calculation involving C$_1$, C$_2$, and D$_1$ (shown in red), we obtained, above 1 eV, an average increase by a factor of 4 or even more. The resonance structure of the cross section shows a more pronounced multi-core character, although some of the resonances have been displaced and broadened, with loss in peak intensity.

The finer details of the cross sections can be seen in the enlarged Figs.~\ref{fig:DRcscoreD1} and \ref{fig:DRcscoreD2}, in which the cross sections were rescaled for better visibility.  One can observe the typical resonant pattern in the cross section, induced by the temporary capture into the vibrational levels of the Rydberg states. The ionization limits (vibrational levels of the ion, represented as dashed vertical lines in the figures, green for the ground core C$_1$ and red for the excited-core C$_2$) act as accumulation-points for the resonances in the cross sections. This resonant structure clearly shows the multi-core effects. In Fig.~\ref{fig:DRcscoreD1}, when going from the green (C$_1$ - D$_1$) curve to the red one (C$_1$ \& C$_2$ - D$_1$) and to the blue one (C$_1$ \& C$_2$ - D$_1$ \&  D$_2$), further resonances appear due to the Rydberg series built on the excited core C$_2$  (marked with red triangles in the figures) as well as new accumulation points  (ionization limits of the excited core, shown by vertical dashed red lines). The same features are visible in Fig.~\ref{fig:DRcscoreD2}, where we compare the green (C$_1$ - D$_2$) curve to the red one (C$_1$ \& C$_2$ - D$_2$) and to the blue one (C$_1$ \& C$_2$ - D$_1$ \&  D$_2$). Figure~\ref{fig:DRcshe} focuses on the contribution of the dissociative state D$_2$ to the {\it total} DR cross section in the high energy range. In accordance with the case of the {\it direct} mechanism, illustrated in Fig.~\ref{fig:DRdircs}, we found that the inclusion of this state (corresponding to the blue curve)  increases the DR cross section by up to one order of magnitude  compared to the case (shown by the red curve) where only D$_1$ was considered.

We have evaluated the Maxwell isotropic rate coefficients, starting from the computed cross section, for a broad range of electronic temperatures, relevant especially for cold non-equilibrium environments. Figure~\ref{fig:DRrate} shows the DR rate coefficients of the vibrationally relaxed SH$^+$ for the already mentioned case studies, for both {\it direct} and {\it total} processes.
The inclusions of the excited core C$_2$ (red curves) and, eventually, of the dissociative state D$_2$ (blue curves), in addition to the  core C$_1$ and the dissociative state D$_1$ (green curves) increase considerably the reaction rate coefficients, in average by a factor of 2. At the same time this figure shows the relevance of the resonant {\it indirect} mechanism in the DR process, putting in evidence their importance in the low and high energy and/or temperature regions.

A more meaningful and clearer comparison with the experiments is obtained by convoluting the total cross sections with the anisotropic Maxwell distribution given in Ref.~\onlinecite{becker2015}. The convolution procedure smooths out the numerous narrow resonances originating in the capture of the incoming electron into high $n$ neutral Rydberg states of ground ionic core C$_1$, but keeps quite visible the broad resonances that occur due to the presence of core excited Rydberg states, i.e. those built on the C$_2$ core. Figure~\ref{fig:DR2coreanirate} shows the result of this convolution for our most elaborate model, which includes two ion cores and two dissociative states (blue line), in comparison with the experimental data measured on the TSR-storage ring~\cite{becker2015}. The agreement of our calculated rate coefficients with the experimental one is satisfactory above 10 meV collision energy and persists up to 1 eV. The broad resonances in the experimental rate can be associated with those induced by the core-excited Rydberg states appearing in the computed cross section, but they are shifted towards lower energy. %
The two rates disagree below 10 meV. Since the very low energy region near 10 meV  that contibutes to the convoluted rate is very rich in constructive and destructive resonances, the rate in this region is extremely sensitive to the molecular structure data. Consequently, small variations in the crossing points between the dissociative states and the ion states, as well as in the quantum defects and couplings, may induce notable variations in the cross section and anistropic rate coefficients. Other electronic states, such as the $^4\Pi$ discussed in Section~\ref{sec:SOCIcalcs}, might also contribute to DR at low energies. Finally, rotational effects, also not accounted for in the present study, may play a non-negligible role~\cite{motapon2014,epee2015}.

\section{Conclusion and future work} \label{sec:conclude}

Using extensive MRCI calculations, we have calculated PECs at the same level of theory for the $^3\Sigma^-$ and $^1\Delta$ states of \SHplus\ and also for the $^2\Pi$ excited valence and Rydberg states of SH. We extracted the diabatic potential energy curves of the two lowest $^2\Pi$ autoionizing states, which dissociate to S($^3P$) and S($^1D$). We also determined the two lowest SH Rydberg states, one with the ground state ($^3\Sigma^-$) ion core and one with the first excited state ($^1\Delta$) ion core. These four states of SH have the most important Rydberg-valence coupling for the DR process. Using the diabatic potential energy curves and electronic couplings  obtained from the block diagonalization method, we have calculated the cross sections and the rate constants for the DR of cold SH$^+$ ($v_i^+ = 0$) with the MQDT method.

The progressive introduction of reaction channels and of their interactions in our treatment allowed us to explicate the role of the valence states and of the ground core and core-excited Rydberg states in the dissociation dynamics. Our Maxwell anisotropic rate coefficient is in satisfactory agreement with the experimental rate coefficient measured in the TSR storage ring in the range 10 meV -- 1 eV. Whereas the present approach takes into account the full vibrational structure of the relevant electronic states of SH$^+$ and SH, assuming complete rotational relaxation, further work will be necessary in order to take into account the role of rotational effects (excitation and interchannel couplings) on the magnitude of the rate coefficient at very low energy.

It is also possible that the low energy rate constants are influenced by electronic PECs and coupling terms of other symmetries. Further electronic structure calculations for other symmetries, such as the $^4\Pi$ and possibly the $^2\Sigma^+$, can now be carried out with confidence to investigate this possibility, since our methodology has been validated for the $^2\Pi$ case.

Another direction for future work will be to consider the branching ratios for producing the $^3P$ and $^1D$ states of S\@.  Calculating these branching ratios will require taking into account the multiple curve crossings involving the Gnd curve and the dissociating curves D$_1$ and D$_2$ at large $R$ (see Fig.~\ref{fig:FOCI_DB}).  Electronic coupling terms necessary to treat these crossings are available from the quantum chemical calculations of \Hdia.  The MQDT formalism is currently restricted to short range interactions and will have to be combined with a Landau-Zener model for the curve crossings.

\section*{Acknowledgments}

A.P.H. acknowledges support by NSF under grant PHY-1403060. Computational facilities at the Texas Advanced Computing Center (TACC) used for this research were supported by NSF through XSEDE resources provided by the XSEDE Science Gateways Program. D.T., I.F.S. and J.Zs.M.  acknowledge support from the French national PCMI program funded by the Conseil National de la Recherche Scientifique (CNRS) and Centre National d'Etudes Spatiales (CNES). D.O.K and O.E.D. acknowledge support from the DoD-HPCMP and from USMA. J.Zs.M and I.F.S acknowledge the French ANR-HYDRIDES grant, the LabEx EMC$^3$ via the project PicoLIBS (No.\ ANR-12-BS05-0011-01), the BIOENGINE project (sponsored by the European fund FEDER and the French CPER), the F\'ed\'eration de Recherche Fusion par Confinement Magn\'etique - ITER and the European COST Program CM1401 ``Our Astrochemical History.''\\

\section*{Data availability}
Upon a reasonable request, the data supporting this article will be provided by the corresponding author.

\bibliographystyle{unsrt}
\bibliography{allbib5,DahbiaBIB,ZsoltBIB}

\end{document}